\documentclass[preprint,article,accept,moreauthors,10pt,a4paper]{mdpi}

\firstpage{1} 
\pubvolume{xx}
\issuenum{1}
\articlenumber{1}
\pubyear{2018}
\copyrightyear{2018}
\history{version 14/08/2018}

\usepackage{pgfplots,xcolor}
  \pgfplotsset{compat=newest}
  \usetikzlibrary{plotmarks}
  \usetikzlibrary{arrows.meta}
  \usepgfplotslibrary{patchplots}
  \usepackage{grffile}
  \usepackage{amsmath,amssymb}
\usepackage{graphicx,placeins,physics}
\usepackage[caption=false]{subfig}
\usepackage{dcolumn}
\usepackage{bm}
\usepackage[normalem]{ulem}

\DeclareMathOperator{\aop}{\hat{a}}
\DeclareMathOperator{\cop}{\hat{a}^\dagger}
\DeclareMathOperator{\Jop}{\hat{J}}

\newcommand{\delop}{\hat{\delta}}

\DeclareMathOperator{\eada}{\expval{\cop \aop}}

\newcommand{\edtwo}{\expval*{\delop\delop}}
\newcommand{\edntwo }{\expval*{\delop^\dagger\delop}}

\newcommand{\unev}[1]{\widetilde{\ev{#1}}} 

\DeclareMathOperator{\xop}{\hat{X}}
\DeclareMathOperator{\pop}{\hat{P}}

\newcommand{\ph}{\theta}
\newcommand{\phop}{\hat{\ph}}

\newcommand{\cphexp}{e^{-i\phop}}
\DeclareMathOperator{\nop}{\hat{n}}
\DeclareMathOperator{\en}{\ev{\nop}}
\DeclareMathOperator{\enn}{\expval*{\nop\nop}}
\newcommand{\dn}{\hat{\delta}_n}
\newcommand{\dph}{\hat{\delta}_{\ph}}
\newcommand{\edndn}{\ev*{\dn\dn}}
\newcommand{\eph}{\ev*{\phop}}
\newcommand{\ephph}{\ev*{\phop\phop}}
\newcommand{\edphdph}{\ev*{\dph\dph}}
\newcommand{\enph}{\ev*{\nop\phop}}
\newcommand{\ephn}{\ev*{\phop\nop}}
\newcommand{\edndphsym}{\ev*{\dn\dph}_{\text{sym}}}
\DeclareMathOperator{\Ham}{\hat{H}}
\newcommand{\Rho}{\hat{\rho}}
\DeclareMathOperator{\Oop}{\hat{O}}

\DeclareMathOperator{\Var}{Var}

\DeclareMathOperator{\intravar}{\Var_1}
\DeclareMathOperator{\intervar}{\Var_2}

\newcommand{\jump}{\xrightarrow{J}}

\newcommand{\new}[1]{#1}

\Title{Gaussian quantum trajectories for the variational simulation of Open Quantum-Optical Systems}

\Author{Wouter Verstraelen and Michiel Wouters}

\AuthorNames{Wouter Verstraelen and Michiel Wouters}

\address{ \quad TQC, Universiteit Antwerpen, Universiteitsplein 1, B-2610 Antwerpen, Belgium}

\corres{Correspondence: wouter.verstraelen@uantwerpen.be (W.V.); michiel.wouters@uantwerpen.be (M.W.); Tel.: +32-3265-2480 (M.W.)}

\abstract{We construct a class of variational methods for the study of open quantum systems based on Gaussian ansatzes for the quantum trajectory formalism. Gaussianity in the conjugate position and momentum quadratures is distinguished from Gaussianity in density and phase. We apply these methods to a driven-dissipative Kerr cavity where we study dephasing and the stationary states throughout the bistability regime.
Computational cost proves to be similar to the truncated Wigner (TWA) method, with at most quadratic scaling in system size. Meanwhile, strong correspondence with the numerically exact trajectory description is maintained so that these methods contain more information on the ensemble constitution than TWA and can be more robust.}


\featuredapplication{nonlinear, driven-dissipative, photonic cavity -- optical bistability and dephasing}

\begin{document}


\section{Introduction}
Many-body systems of interacting photons have come under intense investigation over the last years, with both circuit QED and semiconductor heterostructure based systems \cite{photontransition,photonsimulation,lightsimulation,polaritonsuperfluidity,zeroandpistates,honeycomb}. 
The main difference with traditional many-body systems, both in hard and synthetic condensed matter, is the fact that the photon lifetime is typically shorter than the time at which the system dynamics develop. To compensate for the losses, photonic systems have to be continuously driven, yielding a steady state that is a balance between driving and dissipation.
Even though the basic theoretical framework for the description of open systems is well understood \cite{carmichaelbook,breuer}, the exponential complexity of the many-body problem requires the development of practical approximate techniques to tackle systems that consist of many excitations of multiple modes, stirring the need for methods that unite the approaches of these different fields \cite{daleyreview}.

In order to study the time-evolution of an open quantum system, there are two distinct approaches \cite{breuer}. The first one is the study of the ensemble as a whole. For a Markovian system this translates into the Lindblad master equation for the density matrix $\hat{\rho}$, where $\hat{\rho}$ contains $D^2$ elements for Hilbert-space dimension $D$. The second approach is the trajectory approach: here only a pure wave-function $\ket{\psi}$ of $D$ elements is evolved at a time. Through stochastic decoherence, the environment effectively introduces additional classical noise that complements a non-hermitian Schr\"odinger evolution of the system, and the ensemble evolution is obtained by averaging over many trajectory evolutions. In addition to the reduction of complexity from $D^2$ to $D$, another advantage of the trajectory approach is that the full statistics of detector clicks are obtained.

After preliminary work by Davies \cite{davies1969}, this quantum trajectory formalism was developed towards its current shape \cite{trajectoryreview,carmichaelbook} by a number of different groups \cite{dalibard1992,zoller1992,Carmichaelorig,barchielli1991}. It is also known under a variety of other names: \emph{Monte-Carlo wave function method}, \emph{quantum jump method} or \emph{Stochastic Schr\"odinger equation}. The act of averaging over many such trajectories to obtain an ensemble description is often called the \emph{Stochastic simulation Method}. 
The formalism behind quantum-trajectories is tightly bound to quantum metrology \cite{qmac}.
In this context, is important to note that quantum trajectories are not a single method, but rather a class of methods, depending on the so-called unraveling which corresponds to some (hypothetical) measurement protocol taking place on the environment. For quantum optical systems, the most common protocols are photon-counting, resulting in piecewise-deterministic processes on the one hand and homodyne/\mbox{(far-detuned-)}heterodyne detection resulting in Wiener processes on the other hand \cite{breuer}. The heterodyne case has proven to be equivalent to the stochastic collapse model of quantum state diffusion \cite{breuer}. \new{A summary of these common measurements and their relation with quantum trajectories is given in Appendix \ref{sec:unravelings}.}

Although analytical effort can sometimes provide a serious reduction \cite{prosenexact,fleishhauer}, computational complexity typically remains a limiting factor in exact numerical study of open systems. The aforementioned quantum trajectory method, which allows for a description in the Hilbert space of pure states $\mathcal{H}$ instead of the superspace  of density matrices $\mathcal{H}^2$, can reduce the memory-cost for individual simulations to the level of closed-system problems, at the expense of the need to average over many individual Monte Carlo realizations.
Still, as in the closed system case, many-body systems are usually too complex for exact computation and one needs to use approximate methods instead. These can either occur at the level of the Master equation or at the level of individual trajectories  \cite{daleyreview}.
We are interested in a method that is variational and provides a description on the level of trajectories. 
Some current variational (TDVP, \emph{Time-Dependent Variational Principle} \cite{tdvprev}) approaches to the description of open systems aim to describe the system at the level of the Master equation \cite{weimerPRL,varint,varDattani,multisite,optimisation}, with ansatzes including Gutzwiller Density Matrix-\cite{gutzdm1,gutzdm2,TDVP}, Matrix-Product State \cite{matrixprod,Manzoni2017} and Matrix-Product Operator \cite{savonaMPO,PhysRevLett.114.220601,MPSverstraete,MPSsuperspace} methods.

Variational descriptions on the trajectory level have so far focused towards lattice-like systems: systems where the complexity arises mainly due to a large amount of modes while the Hilbert space per mode remains small, with methods such as Gutzwiller Monte Carlo \cite{gutzmcwim1,gutzmcWim,gutzmczoller} or based on t-DMRG \cite{tdmrg1,tdmrg2,tdmrg3}.

Somewhat surprisingly, except for some works on the back-reaction of measurements on quantum many-body systems \cite{Mekhov1,Mekhov2} , the class of Gaussian states \cite{quantumnoise} has not received much attention as a variational ansatz for the simulation of quantum trajectories. This stands in stark contrast with equilibrium systems, where the Gaussian ansatz has been fruitfully exploited, regarding Hartree-Fock-Bogoliubov \cite{hfb,hfbbosons}, and with the important application of time-dependent BCS/BdG \cite{bdgbook,castinBCSstochastic} theory.
It is the purpose of this work to develop the formalism of a Gaussian variational description of quantum trajectories.

\new{Since the Wigner distribution of a Gaussian state is entirely positive, this description is expected to work well in the regime of many particles and weak interactions, when the Wigner distribution is positive everywhere \cite{HUDSON1974}}. In this regime, a different stochastic technique, the so-called truncated Wigner approximation (TWA) is widely used \new{\cite{qfloflight,tworigSAL,tworigSAL2,POLKOVNIKOV20101790,twcons,TWAadd1,TWAadd2,TWAadd3}}, which fits in a broader context of phase-space methods \new{\cite{quantumnoise,qfloflight,TWAexperiment1}}. Here, the starting point is the Fokker-Planck equation for the Wigner function of the system. If one neglects derivatives of order higher than two, this Wigner function can be sampled through Feynman-Kac onto stochastic differential equations (an alternative formulation for open systems is given in \cite{hebensteit}). As opposed to the aforementioned trajectory approach, single samples of TWA do not correspond to a physical state but rather sample the Wigner phase-space. Though TWA can yield very good results, it is not always well-controlled and may predict unphysical behavior \cite{belland}. 

With this work, we wish to close the gap between the robustness of exact quantum trajectory methods and the numerical efficiency of the TWA method.
It is organized as follows: in the next section \ref{sec:trajcorrgaus}, we refer to the quantum trajectory method, how it affects expectation values and the meaning of a Gaussian ansatz. In \ref{sec:kerrXP} we translate this idea to the states that are commonly known as `Gaussian states', to which we will refer for clarity as `$XP$-Gaussian states'. We will apply such an $XP$-Gaussian trajectory method to study the stationary state of a driven-dissipative Kerr-cavity in the bistability regime \cite{DrummondWalls}. Typical Bogoliubov expansion around the mean-field value is fundamentally unable to describe the hopping between the two branches. We will see that these $XP$-Gaussian methods on the other hand can make reasonable predictions of the exact solution. 
In section \ref{sec:phdiffNT} however, we will see that $XP$-Gaussian methods are unable to describe situations where a large variance in phase is present. A convenient alternative that does succeed in providing accurate descriptions are $N\Theta$-Gaussian methods, meaning that the state is Gaussian in density and phase. In section \ref{sec:comp} we will adress computational issues and compare performance with TWA and exact trajectory methods. Finally, in section \ref{sec:concl} we conclude the work. Throughout our work, we will investigate the effect of different unraveling schemes (photon counting and heterodyne detection) on the accuracy of our variational method.

\section{Quantum trajectories for expectation values and Gaussianity} \label{sec:trajcorrgaus}
\subsection{Quantum trajectories for expectation values}

For the remainder of this work, we assume for simplicity systems where the interaction with the environment is entirely characterized by their Markovian dissipation (photon-leaking towards empty space) at rate $\gamma$. At the level of the master equation for the ensemble density matrix $\Rho$, such a system with Hamiltonian $\Ham$ is described by
\begin{equation}\label{eq:lindblad}
\partial_t \Rho=-i\comm{\Ham}{\Rho}+\frac{\gamma}{2}\left(2\aop\Rho\cop-\cop\aop\Rho-\Rho\cop\aop\right),
\end{equation}
with $\aop (\cop)$ the photonic annihilation(creation) operators.

\subsubsection{Photon-counting unraveling}

According to the quantum trajectory method, this master equation \eqref{eq:lindblad} is unraveled into the evolution of stochastic wavefunctions by performing continuous measurement on the environment. The simplest unraveling is photon-counting (\emph{PC}) detection: in between detector-clicks, the wavefunction is propagated as 

\begin{equation}\label{eq:intrtraj}
i\partial_t\widetilde{\ket{\psi}}=\left(\Ham-i\frac{\gamma}{2}\cop\aop\right)\widetilde{\ket{\psi}},
\end{equation}
where the tilde \new{notation ($\widetilde{\,\cdot\,}$)} denotes that the norm of the wavefunction is not conserved. The detection of a photon corresponds to a discrete jump
\begin{equation}\label{eq:jump}
\widetilde{\ket{\psi}}\rightarrow\aop\widetilde{\ket{\psi}}/\norm{\aop\widetilde{\ket{\psi}}}
\end{equation}
when the norm has decreased to  $\norm{\widetilde{\ket{\psi}}}^2=R$, where the random number $R$ has a uniform distribution on the interval $[0,1]$.

In terms of \new{unnormalized} expectation values, equation \eqref{eq:intrtraj} corresponds to the evolution
\begin{align}\label{eq:pcunnorm}
d \unev{\Oop}=&i\unev{\comm{\Ham}{\Oop}}dt-\frac{\gamma}{2}\unev{\cop\aop \Oop}dt-\frac{\gamma}{2}\unev{\Oop\cop\aop}dt.
\end{align}
By choosing $\Oop=1$, we see that the evolution for the normalized expectation value $\ev{\Oop}=\unev{\Oop}/\unev{1}$ is given by
\begin{align}\label{eq:pcnorm}
d \ev{\Oop}=&i\ev{\comm{\Ham}{\Oop}}dt\nonumber\\&-\frac{\gamma}{2}\ev{\cop\aop \Oop}dt-\frac{\gamma}{2}\ev{\Oop\cop\aop}dt+\gamma\eada\ev{\Oop}dt.
\end{align}
Meanwhile, according to \eqref{eq:jump}, a jump affects the expectation value $\ev{\Oop}$ by
\begin{equation}\label{eq:corrjump}
\ev{\Oop}\jump\frac{\unev{\cop\Oop\aop}}{\unev{\cop\aop}}=\frac{\ev{\cop\Oop\aop}}{\ev{\cop\aop}}.
\end{equation}
equations of the form \eqref{eq:pcnorm} and \eqref{eq:corrjump} together then in principle provide a complete description of the evolution of expectation values for a trajectory under photon-counting unraveling.

\subsubsection{Homo- and heterodyne unravelings}

Apart from photon-counting, (generalized) homodyne detection can be performed, where the emitted light is interfered with a classical reference signal (local oscillator) $\beta=\abs{\beta}e^{i\omega_{LO}t}$, resulting in the measurement of the quadrature variables $X$ and $P$ \footnote{\new{they are defined by $\hat{X}=\frac{\aop+\cop}{2}$ and $\hat{P}=\frac{\aop-\cop}{2i}$, analogous to the position and momentum variables of a harmonic oscillator \cite{GaussQInf}. }}. Homodyne measurement corresponds to jump operators $\Jop=\sqrt{\gamma}(\aop+\beta)$. One readily finds that by inverting this relation towards $\aop=\frac{\Jop}{\sqrt{\gamma}}-\beta$ and substituting in \eqref{eq:lindblad}, again an equation in the Lindblad form is retrieved when absorbing a contribution $\sqrt{\gamma}\Im[\beta^*\Jop]$ in the Hamiltonian. Therefore, homodyne detection is equally valid as an unraveling of the Master equation. In practice, because $\beta$ is macroscopic, a diffusion approximation of the jumps is justified \cite{breuer}, yielding the It\^{o} equation

\begin{align}\label{eq:homexacttraj}
d\widetilde{\ket{\psi}}=\left(-i\Ham dt-\frac{\gamma}{2}\cop\aop dt+\gamma\ev{\cop}\aop dt
+\sqrt{\gamma}\aop dZ^*\right)\ket{\psi}
\end{align}
for the unnormalized wavefunction $\widetilde{\ket{\psi}}$. 
Here we have divided the decay rate in two channels $\gamma=\gamma_X+\gamma_P$ for which either $X$ or $P$ are monitored. That is, if $\beta$ is in-phase with the cavity field, there is only one independent measurement record and we speak of `true' homodyne detection, for example if only $X$ is measured (\emph{Hom. (X)}), $\gamma_X=\gamma$ and $\gamma_P=0$. If $\beta$ is far-detuned from the cavity field on the other hand, both quadratures are equally measured such that $\gamma_X=\gamma_P=\frac{\gamma}{2}$. This latter detection scheme is known as heterodyne detection (\emph{Het.}). From the next section on, we will assume this heterodyne case unless indicated otherwise.
The complex Wiener noise in Eq. \eqref{eq:homexacttraj} has a Gaussian distribution with zero mean and variance $dZ=\sqrt{\frac{\gamma_X}{\gamma}}dW_X+i\sqrt{\frac{\gamma_P}{\gamma}}dW_P$ with  $dW_X^2=dW_P^2=dt$ and $dW_X dW_P=0$, meaning that $\abs{dZ}^2=dt$ and, in the heterodyne case, $dZ^2=dZ^{*2}=0$.

Analogously to the deterministic part of photon-counting, \eqref{eq:homexacttraj} can be translated to the evolution of normalized expectation values, yielding 
\begin{align} \label{eq:homgen}
d\ev{\Oop}=&i\ev{\comm{\Ham}{\Oop}}dt\nonumber\\&-\frac{\gamma}{2}\ev{\nop \Oop}dt-\frac{\gamma}{2}\ev{\Oop\nop}dt+\gamma\ev{\cop\Oop\aop}dt\nonumber\\
&+\sqrt{\gamma}\left(\ev{\cop\hat{\delta}_O}\,dZ+\ev{\hat{\delta}_O\aop}\,dZ^*\right),
\end{align}
where we have introduced the generic fluctuation 
${\hat{\delta}_O=\Oop-\ev{\Oop}}$.

\subsection{Closing at Gaussian level}

Equations \eqref{eq:pcnorm} or \eqref{eq:homgen} will generally introduce an infinite hierarchy of correlation functions. This is a similar situation to classical mechanics, where the Liouville equation introduces the BBGKY hierarchy, that must be truncated at some point. At the level of evolution for the full ensemble (the master equation), a systematic derivation of proper truncation schemes (also known as cumulant expansions) for driven-dissipative quantum systems has been performed in \cite{wimcorrfties}. Generally, one expects the resulting predictions to converge as more correlation functions of higher order are taken into account. Nevertheless, it is only at the mean-field or at the Gaussian level that it is possible to \emph{close} the hierarchy instead of truncating, ensuring that the state remains physical by construction. When closing \eqref{eq:pcnorm}, \eqref{eq:homgen} at mean-field level, the dynamics are trivial and coincide with the mean-field solution of the master equation. At the Gaussian level on the other hand, the closing of the equations can be performed by applying Wick's theorem and does reflect the stochasticity of the unravelings.

As an ansatz, we thus approximate the state to be contained in a Gaussian subspace of the Hilbert space. These Gaussian states are entirely characterized by their first and second moments, keeping the amount of independent variables of which the evolution needs to be studied limited. This provides a large reduction of complexity from the whole Hilbert-space, which scales exponentially with system size (the number of modes).
Because of the presence of the second moments, a Gaussian ansatz is by nature one order higher than the mean-field approach that corresponds to a coherent ansatz. Formally then, the Gaussian methods are of the same order as Bogoliubov theory. Nevertheless, we can expect these Gaussian trajectory methods to provide a more refined description of the whole state because the second-order expansion is done at the level of individual trajectories instead of on the level of the ensemble. In the next section, we will illustrate this with the example of bistability in a Kerr nonlinear cavity. Where the Bogoliubov theory is only able to describe the fluctuations around one of the two stable branches, the variational Gaussian method is also capable of describing the switching between the branches. 

\new{The fact that our method is variational can be seen from the following argument. By using the Wick theorem, we clearly restrict to the manifold of Gaussian states. The variational principle consists of evolving the state (by which it will leave the manifold) and then projecting is back.
For such a state, the time evolution of the first and second order correlators over a small time interval can be computed by applying Wick’s theorem to the right hand side of the Heisenberg equation. 
The next step is to project back to the Gaussian manifold. This projected state will have to a good approximation the same first and second order correlation functions as the evolved state. }

\section{Kerr-bistability and the \texorpdfstring{$XP$}{XP}-Gaussian methods}\label{sec:kerrXP}

As a first example, we look at a driven-dissipative cavity with Kerr non-linearity, described by the Hamiltonian

\begin{equation}\label{eq:ham}
\Ham=-\Delta \cop\aop+\frac{U}{2}\cop\cop\aop\aop+Fa^{\dagger}+F^{*}\aop
\end{equation}
where $\Delta$ the cavity-laser detuning, $U$ the photon-photon interaction and $F$ the classical laser amplitude. Furthermore, photons leak to the vacuum with rate $\gamma$ as described in section \ref{sec:trajcorrgaus}.

Exact solutions for the expectation values $\ev{\cop^m\aop^n}$ of the stationary state, as well as a semi-classical model, were calculated in \cite{DrummondWalls}. Interestingly, at the classical level, this system features a bistability regime where, given all other parameters, two stable solutions for the density exist.

Such a Kerr-cavity provides a model for polariton condensates \cite{qfloflight}, for which it has been shown that trajectory methods can provide an adequate description \cite{Wouters2012,Kuznetsov2017}. Also coupled arrays of these cavities are an object of current interest \cite{aleboite} and it is in these systems that the TWA method has been proven insufficient \cite{belland}.

As it is the most straightforward ansatz corresponding with the Gaussian approximation of section \ref{sec:trajcorrgaus}, we assume the the state to be Gaussian in the quadrature operators $\xop$ and $\pop$ or, equivalently, $\aop$ and $\cop$, i.e. the states that are colloquially known as Gaussian states \cite{GaussQInf}. Such an $XP$-Gaussian state is characterized entirely by its mean-field value $\alpha=\ev{\aop}$ and its two-point correlation functions $\ev{\aop\aop}$ and $\ev{\cop\aop}$ or, equivalently, $\alpha$, $\edtwo$ and $\edntwo $ where ${\delop=\hat{\delta}_{\aop}}\new{=\aop-\alpha}$. The connected correlation functions equal ${\edtwo=\ev{\aop\aop}-\alpha^2}$ and ${\edntwo =\ev{\cop\aop}-\abs{\alpha}^2}$. 

From \eqref{eq:pcnorm} and after applying Wick's theorem to close the set of equations, we obtain for photon-counting the deterministic evolution

\begin{align}
\partial_t\alpha&=\left(\frac{-\gamma}{2}+i\Delta\right)\alpha-Ui\left(\abs{\alpha}^2\alpha+2\alpha\edntwo +\alpha^*\edtwo\right)-iF-\gamma\left(\alpha\edntwo +\alpha^*\edtwo\right)\\
\partial_t\edtwo&=(2i\Delta-\gamma)\edtwo-Ui\left(\alpha^2(1+2\edntwo )+\edtwo(1+4\abs{\alpha}^2+6\edntwo )\right)-2\gamma\edntwo \edtwo\\
\partial_t\edntwo &=2U\Im\left[\alpha^2\edtwo^*\right]-\gamma\left(\edntwo +\edntwo ^2+\abs{\edtwo}^2\right).
\end{align}

This deterministic evolution is propagated until the norm, evolving through \eqref{eq:pcunnorm}, becomes $\unev{1}=R$. Then, according to \eqref{eq:corrjump}, jumps occur as

\begin{align}
\alpha&\jump\frac{\abs{\alpha}^2\alpha+2\alpha\edntwo +\alpha^*\edtwo}{\abs{\alpha}^2+\edntwo }\\
\edtwo&\jump\frac{\abs{\alpha}^4\edtwo+2\abs{\alpha}^2\edntwo \edtwo-\alpha^2\edntwo ^2+3\edntwo ^2\edtwo-\alpha^{*2}\edtwo^2}{\abs{\alpha}^4+2\abs{\alpha}^2\edntwo +\edntwo ^2}\label{eq:pcdetdd}\\
\edntwo &\jump\frac{\abs{\alpha}^4\edntwo -2\Re\left[\alpha^2\edtwo^*\edntwo \right]+\edntwo \abs{\edtwo}^2+2\abs{\alpha}^2\edntwo ^2+2\edntwo ^3}{\abs{\alpha}^4+2\abs{\alpha}^2\edntwo +\edntwo ^2}.\label{eq:pcdetdnd}
\end{align}

In the heterodyne unraveling on the other hand, the evolution of the mean-field \eqref{eq:homgen} is given by

\begin{align}
d\alpha =&\left[\left(\frac{-\gamma}{2}+i\Delta\right)\alpha-Ui\left(|\alpha|^2\alpha+2\alpha\edntwo +\alpha^*\edtwo\right)-iF\right]dt+\sqrt{\gamma}\left(\edntwo dZ+\edtwo dZ^*\right).
\end{align}

Generally, one would expect similar stochastic equations for $\edtwo$ and $\edntwo $. Remarkably, they are entirely deterministic and, moreover, are identical to the deterministic part of photon-counting equations \eqref{eq:pcdetdd}, \eqref{eq:pcdetdnd}. It should be noted, however, that the the equations of motion are different for more general homodyne detection schemes.

For all unravelings, as a slightly more efficient and stable alternative to evolving $\edntwo $ along explicitly trough \eqref{eq:pcdetdnd}, one can obtain it from $\edtwo$ by asserting that the state remains pure, corresponding to the condition \cite{GaussQInf}
\begin{equation}\label{eq:purityrel}
\edntwo +\edntwo ^2=\abs{\edtwo}^2.
\end{equation}
For an initially pure state, relation \eqref{eq:purityrel} remains exactly satisfied through heterodyne measurement as well as through the deterministic evolution between photon jumps, although the purity
\begin{equation}
\tr[\Rho^2]=\left(1+4\left(\edntwo +\edntwo ^2-\abs{\edtwo}\right)\right)^{-1/2}
\end{equation}
briefly decreases at a jump after which it tends to relax back to 1. Relation \eqref{eq:purityrel} remains fulfilled at a jump up to order $\abs{\alpha}^{-5}$ so that only in systems where there is a significant jump rate at zero density it is better to evolve $\edntwo $ explicitly with Eq. \eqref{eq:pcdetdnd}.

\begin{figure}
\centering
\includegraphics[width=0.80\linewidth]{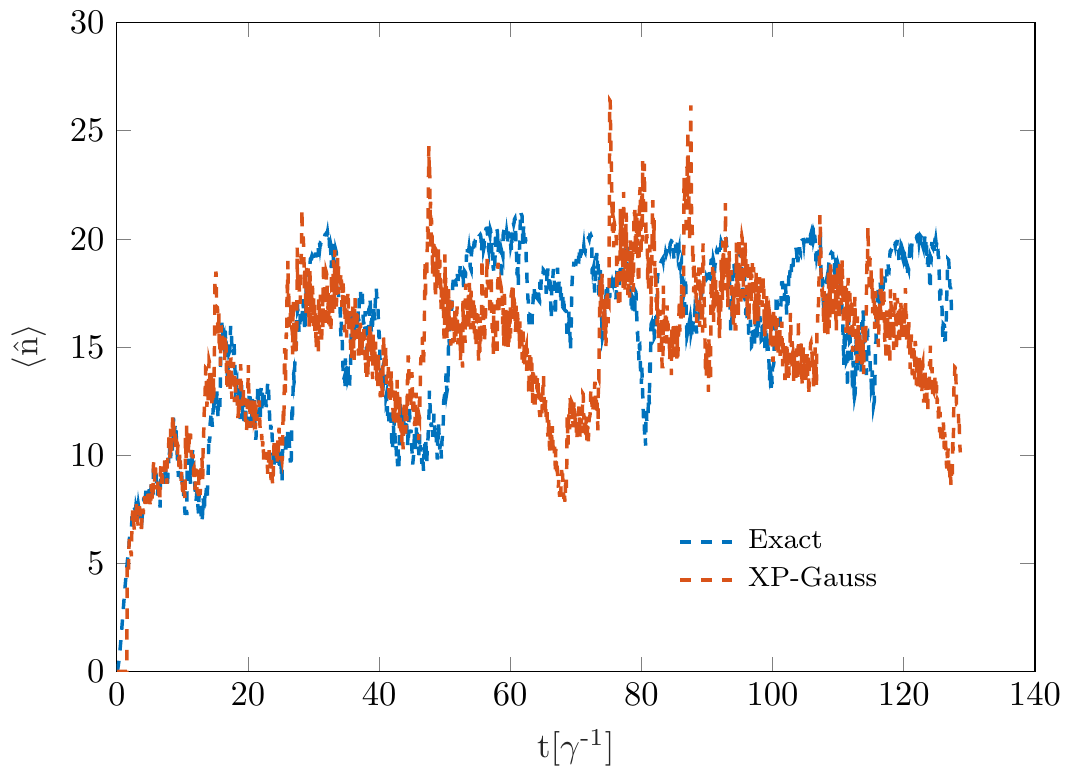}
\caption{\new{Numerically} exact trajectory \new{(up to particle number truncation)} with its corresponding $XP$-Gaussian approximation for a photon-counting process with parameters $U/\gamma=0.05 , \;\Delta/\gamma=1$ and $F/ \gamma =2.235$. }
\label{fig:traject}
\end{figure}

In Fig. \ref{fig:traject}, we show a comparison between single trajectories that were obtained with the numerically exact evolution of the wave function and the Gaussian variational ansatz where for the jumps in both simulations identical random numbers were used.
Initially there is a very strong correspondence with the $XP$-Gaussian variational method while at later evolution times an offset in time emerges and the good correspondence is lost. Qualitatively though, the behavior remains similar so that approximate correspondence in the weak sense (for the whole ensemble) can remain fulfilled.

\begin{figure}

\includegraphics[width=\linewidth]{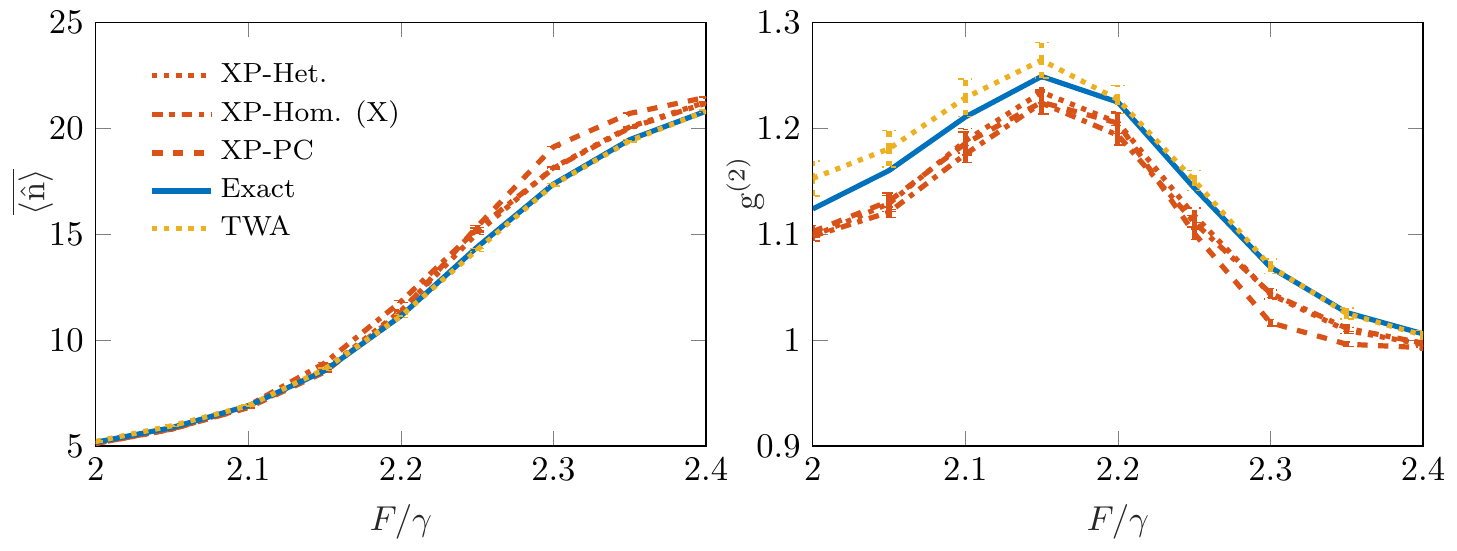}

\caption{Stationary ensemble expectations of photon number (left) and second-order correlation (right) as function of $F$ through the bistability regime for parameters $U/\gamma=0.05,\Delta/\gamma=1$. Results were obtained by averaging over $10^4$ samples after $t=100\gamma^{-1}$ of evolution from the vacuum. \new{The exact solution is the analytical result from \cite{DrummondWalls}.}
} 
\label{fig:bistability}
\end{figure}

To verify this, in figure \ref{fig:bistability} densities and density correlations $g^{(2)}=\langle\cop\cop\aop\aop\rangle/\ev{\cop\aop}^2$ are plotted as function of $F$ throughout the bistability regime of the Kerr-model. We see that the $XP$-Gaussian methods can provide reasonable to very good predictions of the exact correlation functions from \cite{DrummondWalls}. In this regime of relatively low density these $XP$-Gaussian methods are mostly outperformed by the TWA method, though. It is also seen that the predictions on the statistics of the ensemble as a whole depend (weakly) on the choice of unraveling, whereas this is independent for exact trajectories. We should note that, because of the low density, simulation with exact trajectories can also still easily be performed. It is the opposite, high-density, limit where Gaussian ansatzes will provide a better description as well as where exact trajectories become computationally unfeasible.

\section{Phase diffusion and the \texorpdfstring{$N\Theta$}{NTheta}-Gaussian method}\label{sec:phdiffNT}

\subsection{Phase space evolution \label{sec:phspev}}

As another example, we look at the time-evolution of a freely evolving Kerr cavity: we envision an initial state present in system \eqref{eq:ham} where at $t=0$ the pump is turned off, i.e. $F=F_{on}\Theta(-t)$. Without the pump, $\mathcal{U}(1)$ symmetry is restored, allowing the phase to diffuse freely. 

\begin{figure}

\includegraphics[width=\linewidth]{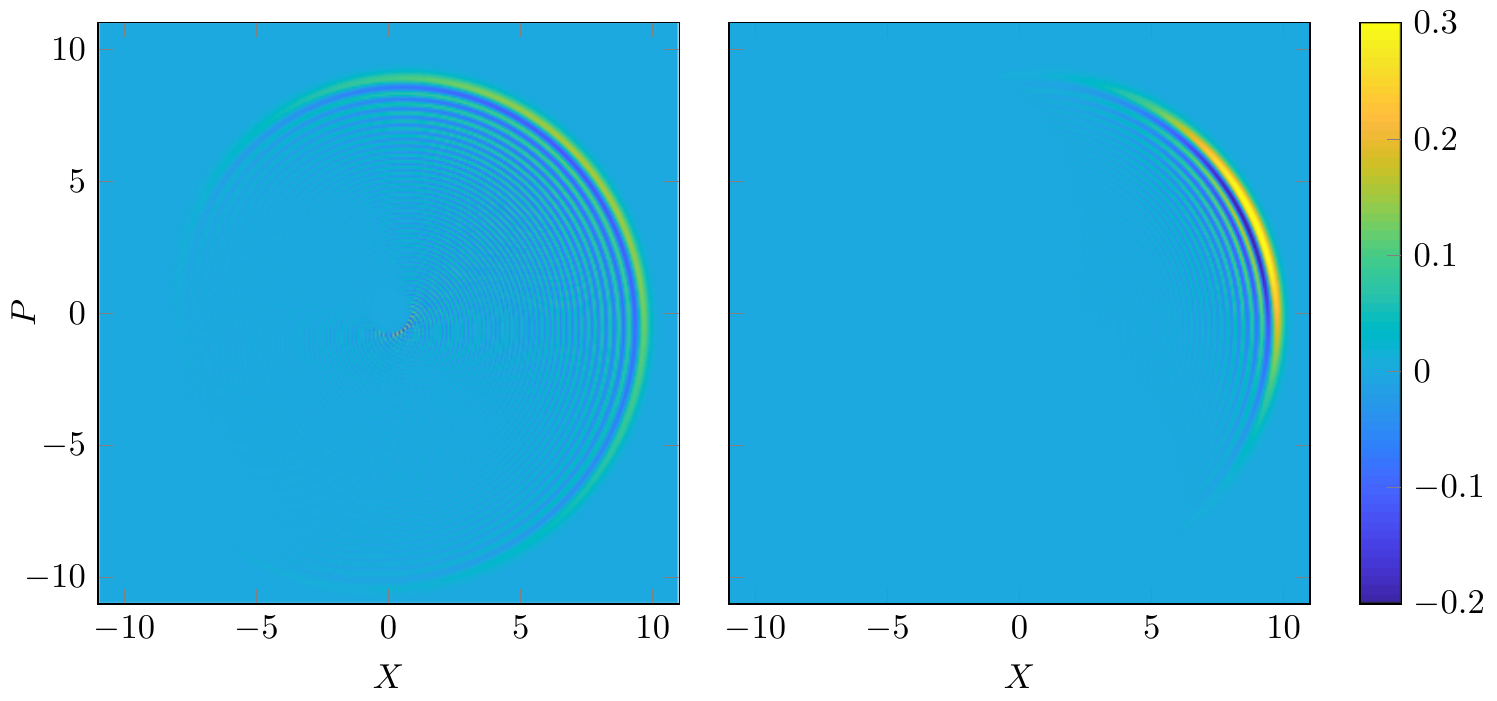}
\caption{Wigner function snapshots of a typical single exact trajectory \new{(obtained by numerical integration in truncated Fock space)} after t=0.1$\gamma^{-1}$ phase diffusion out of an initial coherent $\ket{\alpha}=\ket{10}$ state ($F/\gamma=0,\quad U/\gamma=1,\quad \Delta/\gamma=100$), for photon-counting (left) and heterodyne detection (right). It is clear to see that these states are not $XP$-Gaussian, firstly because they are bended and secondly because the $W$-function exhibits Fock-like negative parts. Note that the intra-sample variance $\intravar$ of the phase is smaller for a heterodyne sample, its inter-sample $\intervar$ variance is larger. Therefore, heterodyne detection is slightly less problematic for the $XP-$Gaussian states.}
\label{fig:Wfties}
\end{figure}

The challenging nature of this problem for our Gaussian variational ansatz can be appreciated from the Wigner distributions of a single realization after some time, shown in Fig. \ref{fig:Wfties}. Panel (a) shows that the phase space distribution has almost spread out over a full circle, implying the loss of phase coherence. The difference between the left and right hand panels, obtained with photon counting and heterodyne detection respectively, shows how the phase space distribution is kept more concentrated under heterodyne measurement as compared to the photon counting. This is expected, because only the heterodyne measurement gives phase information. 

In this example, we clearly see the importance of the unraveling on the applicability of a Gaussian approximation. Unfortunately for the parameters of Fig. \ref{fig:Wfties}, even for the heterodyne measurement, a Gaussian approximation is very crude. The applicability of the Gaussian method is directly related to the intra sample variance of the phase. In general, in a quantum trajectory simulation, the variance of an observable can be written as
\begin{equation}
\Var{(\Oop)}=\intravar{(\Oop)}+\intervar{(\Oop)},
\label{eq:vartot}
\end{equation}
where the intra and inter trajectory variances are respectively defined as \cite{breuer}
\begin{align}
\intravar(\Oop)&=\frac{1}{N_{\text{traj}}}\sum_\alpha\left[\ev{\Oop^2}_\alpha-\ev{\Oop}_{\alpha}^2\right], \label{eq:var1}\\
\intervar(\Oop)&=\sum_\alpha\frac{\ev{\Oop}_\alpha^2}{N_{\text{traj}}}-\left[\sum_\alpha\frac{\ev{\Oop}_\alpha}{N_{\text{traj}}}\right]^2 \label{eq:var2},\\
\end{align}
where index $\alpha$ labels the trajectories.
Only $\Var(\Oop)$ can be measured without unraveling the dynamics and it is the only one that is accessible within the master equation description (and hence the TWA).

The parameter regime where the width of the phase space distribution is expected to become small can be found by requiring that the phase diffusion rate is much smaller than the rate at which phase information is obtained: $U \sqrt{\eada} \ll \gamma \eada $. For the parameters of Fig. \ref{fig:Wfties}, we have $U /(\sqrt{\eada} \gamma) = 0.1 \ll 1$, but still the phase space distribution cannot be accurately approximated by a Gaussian.

\begin{figure}

\includegraphics[width=\linewidth]{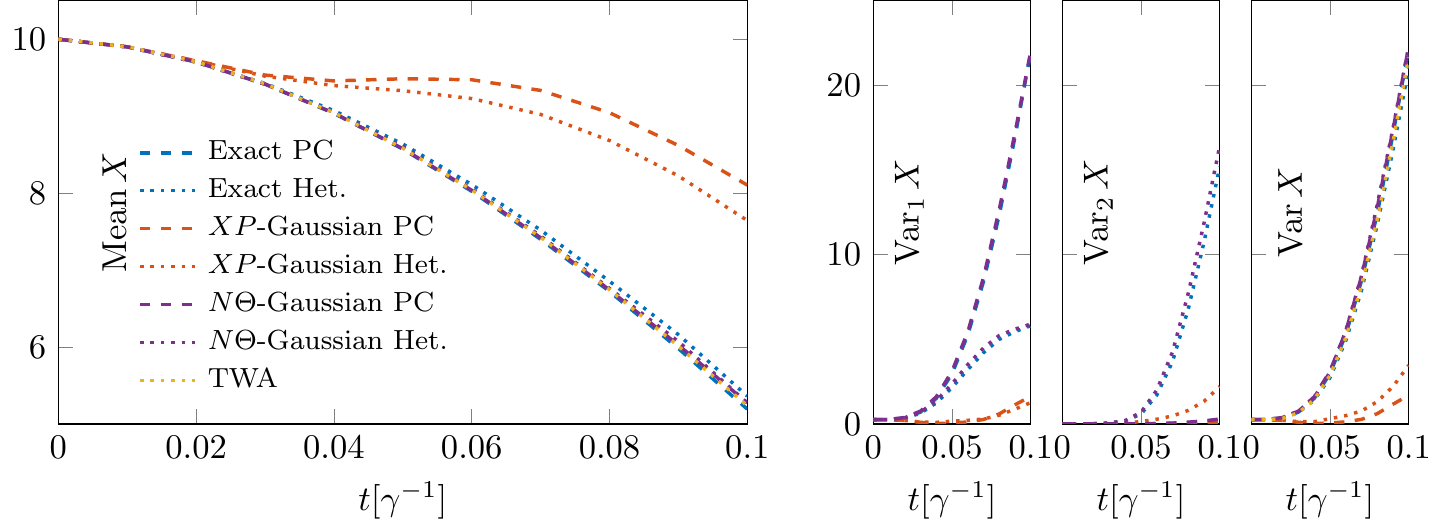}
\caption{
Mean (left panel) and inner-sample-, between-sample- and total variances  (right panel) of the $X$-quadrature obtained by the \new{numerically} exact trajectory methods \new{(in a truncated Fock space)} and TWA as well as $XP$-Gaussian and $N\Theta$-Gaussian variational methods as function of time after free evolution from a coherent initial state $\ket{\alpha}=\ket{10}$ (same parameters as Fig. \ref{fig:Wfties}). Averages are taken over 1e3 samples (1e4 for TWA).  It is clear that the $N\Theta$ variational method provides an accurate description of the true dynamics, both on the level of the whole ensemble as on the level of its constitution in pure states, whereas the latter cannot be obtained from the TWA method.
Very similar behavior is present regarding $P$ and its variances as well as for the $X,P-$covariances. 
}
\label{fig:dephasing}
\end{figure}

\subsection{\texorpdfstring{$N\Theta$}{NTheta}-Gaussian states}

The shape of the Wigner distribution suggests that a Gaussian description in terms of density and phase may provide a better approximation to the quantum trajectory wave functions. It must be noted that strictly speaking, a well-defined hermitian phase operator does not exist \cite{phaseopreview}. Nevertheless, we will work with the Dirac definition $\aop=:e^{i\phop}\sqrt{\nop}$, where $\phop$ defines the phase and $\nop$ is the familiar particle number operator. The practical use of this phase is paramount in quantum hydrodynamics \cite{BEC}. This realm of use coincides with our approximation of density and phase to be continuous and unbounded operators, justified when the density is sufficiently high and the state is localized in phase-space, where the second demand allows avoiding complications arising from the multivaluedness of phase. 

Density and phase are conjugate variables: $\comm{\nop}{\phop}=i$. Assuming the state to be Gaussian in $\nop$ and $\phop$, 
the independent (real) expectation values to take into account are
$\en,\; {\edndn=\enn-\en^2}, \; \eph, \; {\edphdph=\ephph-\eph^2}$ and ${\edndphsym=\enph/2+\ephn/2-\en\eph}$. From these correlation functions, expectation values of products of annihilation and creation operators can be approximately computed by expanding the exponential and using Wick's theorem. For example,
\begin{align}
\ev{\sqrt{\nop}\cphexp} \approx&\sqrt{\en}e^{-i\eph-\frac{\edphdph}{2}}\\&\times\left(1+\frac{1}{2\en}(\frac{1}{2}-i\edndphsym)-\frac{\edndn}{8\en^2}\right).\nonumber
\end{align}
Here, phase-phase correlations were kept up to all orders, but phase-density correlations were truncated at second order. This approximation is again valid when the average particle number is sufficiently large.

Using the given set of expectation values, derivation of the stochastic equations of motion, similarly to the case of $XP$-Gaussian states, proceeds as described in section \ref{sec:trajcorrgaus}. 
For example, for the deterministic evolution under photon-counting we obtain for the density
\begin{equation}
\partial_t \en =-\gamma \edndn +2\Im\left[F \ev{\sqrt{\nop}\cphexp} \right]
\end{equation}
 while jumps
\begin{equation}
\en \jump \en-1+\frac{\edndn}{\en}
\end{equation}
occur when $\unev{1}=R$.
The full set of equations of motion can be found in appendix \ref{sec:NTequations}, with an analytical solution for the $F=0$ case. \new{From a different starting point, coupled equations for the evolution of density, phase and their variances have also been used in \cite{dpcorr1dcondensate}.}

The update rules for the quantum jumps are all exact except the one for $\edphdph$ where an expansion in orders of $\hat n^{-1}$ must be performed. However, in case of a pure state (or more generally constant purity) it can be omitted and, similar to the $XP$-Gaussian case, computed from the other expectation values by the relation
\begin{equation}\label{eq:ptydensphase}
\edndn\edphdph-\edndphsym^2=\frac{1}{4}
\end{equation}
which is the equivalent of \eqref{eq:purityrel}. For a state that is initially pure, relation \eqref{eq:ptydensphase} again remains satisfied up to order $\en^{-5}$ after a jump.

Equations of motion for heterodyne measurement of density-phase Gaussian states are also given in appendix \ref{sec:NTequations}, for example for the density, 
\begin{align}\label{eq:NPpurityrel}
d\en=&\left[2\Im\left[F\ev{\sqrt{\nop}\cphexp}\right]-\gamma\en\right]dt\nonumber\\&+2\Re\left[(\ev{\dn\sqrt{\nop}\cphexp}-\ev{\sqrt{\nop}\cphexp})\sqrt{\gamma} dZ\right]\nonumber\\
\end{align}
is obtained.

On figure \ref{fig:dephasing} (a) the expected evolution of quadrature variable $\hat{X}$ is shown. The right panels show the intra-sample variance $\intravar$, inter-sample variance $\intervar$ and total variances for the evolution of a state that was originally coherent. We see here that the $XP$-Gaussian methods for both the photon-counting and the heterodyne unraveling strongly underestimate the phase diffusion, the description of which is worse for the photon-counting description.

Only in regimes where the importance of losses is much higher than the importance of dephasing can $XP$-Gaussian states provide a suitable description. 
On the other hand, we see that the $N\Theta$-methods are able to capture the phase diffusion on the level of the ensemble as well as TWA. \new{This relative success of the $N\Theta$-methods with respect to the $XP$-methods is sowewhat reminiscent of a similar observation for number-phase phase-space methods of monitored quantum systems \cite{NPW1,NPW2}.} What distinguishes the $N\Theta-$ Gaussian method  from TWA however, is that the $N\Theta$-Gaussian method is able to show the composition of the ensemble: it maintains information of individual trajectories, which is lost in TWA (\new{we note that in practice under appropriate conditions, a single TWA sample may still be representative for experimental realizations \cite{TWAexperiment1,TWAexperiment2}}).

\section{Computational aspects}\label{sec:comp}

Because of the few simple update rules (five real variables-four if purity relations are used), the computational cost for evolving a single variational trajectory is of the same order as the cost for a single TWA sample (two real variables) given the same set of parameters. 
For an exact trajectory $D=2N_{\text{levels}}$ real variables must be evolved for the same single mode, where $N_{\text{levels}}$ is the amount of occupation levels considered. 
If the dimension of configuration space $d$ increases, the dimension of Hilbert space correspondingly grows as $D\propto e^d$, so that large simulations easily become limited by computational load.
The dimension of the corresponding phase space on the other hand only grows $\propto d$, making it far more efficient for large systems. 
For our variational methods it is the amount of distinct Gaussian correlation functions that is the relevant quantity of complexity and this scales in principle as $\propto d^2$. In many large systems, only correlations between neighboring sites are important and other ones can be neglected. If this is the case, scaling $\propto d$ is retrieved, as in Gutzwiller- or tensor-network ansatzes. 

So far, we have only discussed the evolution of a single sample. As Gaussian trajectories, like exact trajectories, have a finite spread in Wigner phase-space, they contain more information than a TWA-sample which is just a point. Therefore, it can be expected that less samples are required to obtain an accurate description of the ensemble dynamics. We can estimate the difference in statistics as follows: take a trajectory method (exact or variational) of which $N_{\text{traj}}$ independent trajectories are evolved. 
For the trajectory methods, the statistical uncertainty \cite{Molmer:93} is
\begin{equation}\label{eq:trajvar}
\sigma^2_{\ev{\Oop},\text{traj}}=\frac{\intervar(\Oop)}{N_{\text{traj}}},
\end{equation}
whereas the analogous uncertainty for TWA is
\begin{align}\label{eq:twavar}
\sigma^2_{\ev{O}_\text{sym},\text{TWA}}&=\frac{\Var_{\text{TWA}}(O_{\text{sym}})}{N_{\text{TWA}}}=\frac{\Var(\Oop)}{N_{\text{TWA}}}\nonumber\\&=\frac{\intravar(\Oop)+\intervar(\Oop)}{N_{\text{TWA}}}.
\end{align}
The ratio of $N_{\text{traj}}$ and $N_{\text{TWA}}$ needed for the same precision can be found by equating \eqref{eq:trajvar} and \eqref{eq:twavar} to be
\begin{equation}\label{eq:samplecriterion}
\frac{N_{\text{TWA}}}{N_{\text{traj}}}=1+\frac{\intravar(\Oop)}{\intervar(\Oop)}.
\end{equation}

\begin{figure}
\includegraphics[width=\linewidth]{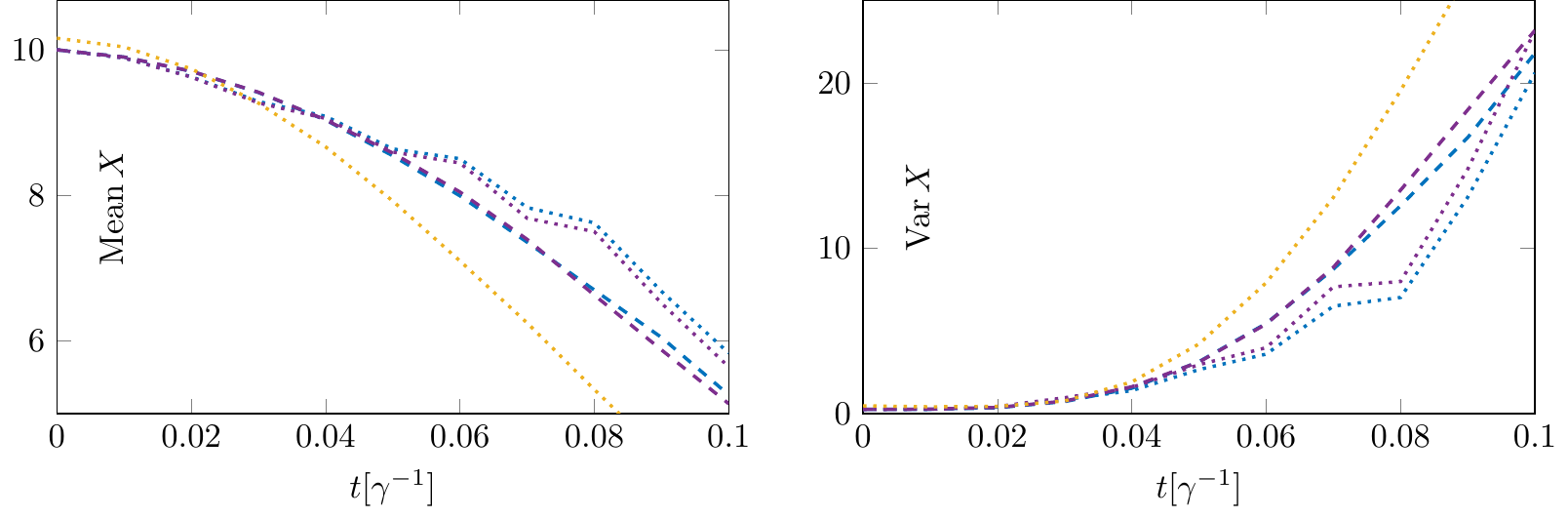}
\caption{Same process as figure \ref{fig:dephasing} (corresponding legend) simulated with ten samples of exact/variational trajectories as well as TWA. 
It is clear that trajectory methods are more tolerant for low statistics than TWA. Corresponding to criterion \eqref{eq:samplecriterion}, we see that photon-counting is still more tolerant than heterodyne detection, corresponding with the fact that $\Var_1(X)\gg\Var_2(X)$ for this photon-counting unraveling, as opposed to heterodyne. Note that the initial state already has a deviation from the true value for TWA but not for the trajectory methods: criterion \eqref{eq:samplecriterion} for the needed ratio of samples diverges there.}
\label{fig:tensamplesonly}
\end{figure}

In this sense, variational trajectories may even outperform TWA computationally. As an example, on figure \ref{fig:tensamplesonly} the same process as on figure \ref{fig:dephasing} is simulated with only ten samples of the exact/variational trajectories and ten samples of TWA. It is clear that TWA provides a bad estimate with such a low number of samples. The performance of trajectory methods with regard to the minimal amount of samples, though observable- and unraveling-dependent, is better. The amount of improvement depends on $\Oop$ and the unraveling: as seen from Eq. \eqref{eq:samplecriterion}, the photon counting unraveling performs best because $\Var_1(X)\gg\Var_2(X)$ as can be appreciated from figure \ref{fig:dephasing}.

When it is a phase-space distribution itself that one is interested in, it can again be challenging to compute it from an exact density matrix or wavefunction, when the particle number is not low \cite{qotoolbox}. If a state is Gaussian, particularly in the $XP$-sense, distributions for individual samples are given by straightforward Gaussian formulas \cite{quantumnoise} that can be added up for the distribution of the whole ensemble. The TWA method is by construction surely also very well suited to plot the Wigner-quasi-probability function, but binning as a histogram is necesarry there, meaning that again a much larger amount of samples is required for a result of high resolution. 

As a final note, all simulations discussed have been performed by trajectories representing pure states. It is also possible to work with trajectories using mixed states, either corresponding to initial classical uncertainty or to imperfect measurements \cite{breuer}. The exact solution of these would require a density matrix, undoing the computational advantage of the stochastic simulation method. For the variational trajectories, the only difference between evolving pure or mixed states is whether constraints \eqref{eq:purityrel} and \eqref{eq:NPpurityrel} are valid or not. Evolving mixed states increases $\intravar(\Oop)$ with respect to $\intervar(\Oop)$ so that less statistics are required according to \eqref{eq:samplecriterion}, at the expense of resolution of the individual trajectories . This would come down to a crossover from pure trajectories to a description on the level of the master equation. Whether such a description is accurate will depend on the system dynamics. For the example of the bistability, if there is not sufficient information on the branch that the system is in, the Gaussian approximation will fail to accurately describe the state, that evolves toward a mixture of the system in the lower and upper branch.
\medskip

\section{\label{sec:concl} Conclusions and Outlook}
We have shown how Gaussian variational quantum trajectory methods can provide a computationally efficient description of open quantum systems. We have explicitly derived the dynamics for both $XP$- and $N\Theta$- Gaussian states, and applied them as an example to a driven-dissipative cavity. $XP$-Gaussian states are always well-defined, though they may be too rigid to describe states with much phase-diffusion. $N\Theta$-states on the other hand, exist only by approximation, but often provide a very good description for the true state as long as the density is sufficiently high. Computationally, the cost of a Gaussian trajectory scales similar to TWA, but typically less samples are needed.  $XP$-Gaussian methods, like TWA, are limited to the semiclassical regime where the Wigner function is always positive. $N\Theta$-Gaussian methods on the other hand can describe the interference patterns similar to Fock-states ($XP$-Gaussian states are the only pure states with an entirely positive Wigner function \cite{GaussQInf}).
As these variational methods\new{, unlike TWA, ensure that the state remains physically well-defined}, we expect them to keep predicting accurate results to a broader class of problems such as coupled cavities. As we have shown, accuracy can be strongly improved by proper choice of the ansatz. Also similar ansatzes, that can for example be Gaussian in other variables \cite{numberanstates}, may be suitable dependent on the problem. As for exact trajectories, the choice of unraveling greatly determines the amount of samples needed for a proper description. In addition, we have seen that this choice of unraveling can also influence the accuracy of the variational method. 
Besides the extension to larger systems, the application to systems with multiple Markovian jump processes is straightforward. Extensions towards more general non-Markovian noise are less clear-cut, but may be feasible through e.g. a doubled Hilbert space framework \cite{breuer}.

\vspace{6pt} 



\funding{This research was funded by the FWO Odysseus program.}

\acknowledgments{We thank Wim Casteels for his important initial contributions to this project; and acknowledge discussion with M. Van Regemortel and D. Huybrechts. Part of the computational resources and services used in this work were provided by the VSC (Flemish Supercomputer Center), funded by the Research Foundation - Flanders (FWO) and the Flemish Government department EWI.}

\conflictsofinterest{The authors declare no conflict of interest.} 

\abbreviations{The following abbreviations are used in this manuscript:\\

\noindent 
\begin{tabular}{@{}ll}
$XP$-Gaussian & state Gaussian in the quadrature variables\\
$N\Theta$-Gaussian & state Gaussian in density and phase\\
TWA & Truncated Wigner Approximation\\
PC & Photon-Counting\\
Hom. (X) & Homodyne detection of $X$-quadrature\\
Het. & Heterodyne detection\\
$\cdot_{\text{sym}}$ & symetrically ordened
\end{tabular}}

\appendixtitles{yes} 
\appendixsections{multiple} 
\appendix

\section{\new{Photon-Counting, homodyne and heterodyne unravelings}}\label{sec:unravelings}
\new{
Dissipative systems, such as a photonic cavity, can be experimentally monitored by continuous weak measurements, typically performed on the photons leaving the cavity. Under such a measurement record, the conditional evolution of the system is known as a \emph{quantum trajectory} \cite{haroche2006exploring} (an elementary introduction to the fundamental principle is given in \cite{simplemodel}). Depending on the nature of the measurement taking place (the \emph{unraveling}), its effect on the state can be through discrete jumps at random times (photon counting) or continuous random noise (real for homodyne detection, complex for heterodyne detection) \cite{walls_milburn,qmac}. An important insight is that the unconditional evolution as described by the dissipative master equation is equivalent to the (classical!) superposition of all possible conditional evolutions. In practice, this justifies the stochastic simulation method where hypothetical trajectories are numerically sampled \cite{haroche2006exploring,carmichaelbook}. We now briefly comment on the nature of the most common unravelings. Measurements are considered idealized, i.e. full efficiency, no dark counts and instantaneous.}
\subsection{\new{Photon Counting}}
\new{The simplest example of an unraveling is photon counting. Here, the detector detects discrete photons leaving the cavity. In every infinitesimal timestep either 0 or 1 photons are detected \cite{walls_milburn}. The detector 'clicks' every time a photon is detected, and in the mathematical descrtiption of the system this is reflected by a jump in the trajectory. However, also the absence of a jump yields information on the photon number in the cavity. For a trajectory this means that the deterministic evolution between the jumps is non-unitary \cite{haroche2006exploring,carmichaelbook}.
Photon counting is an elementary experimental practice to obtain information on photon statistics, including bunching and antibunching \cite{walls_milburn} and has also been proposed as as classroom experiment for students \cite{pcexperiment}.
}
\subsection{\new{Homodyne detection}}
\new{For homodyne detection, the light leaving the cavity is mixed with a strong coherent signal \emph{with the same frequency}, the \emph{Local Oscillator}. (LO) Because of the high intensity, no individual photons are distinguished but a continuous measurement of the photocurrent is performed.	Following the discussion of \cite{scully_zubairy}, the setup is described as follows (see Fig. \ref{fig:homodyne}).
\begin{figure}
\centering
\includegraphics[width=0.75\linewidth]{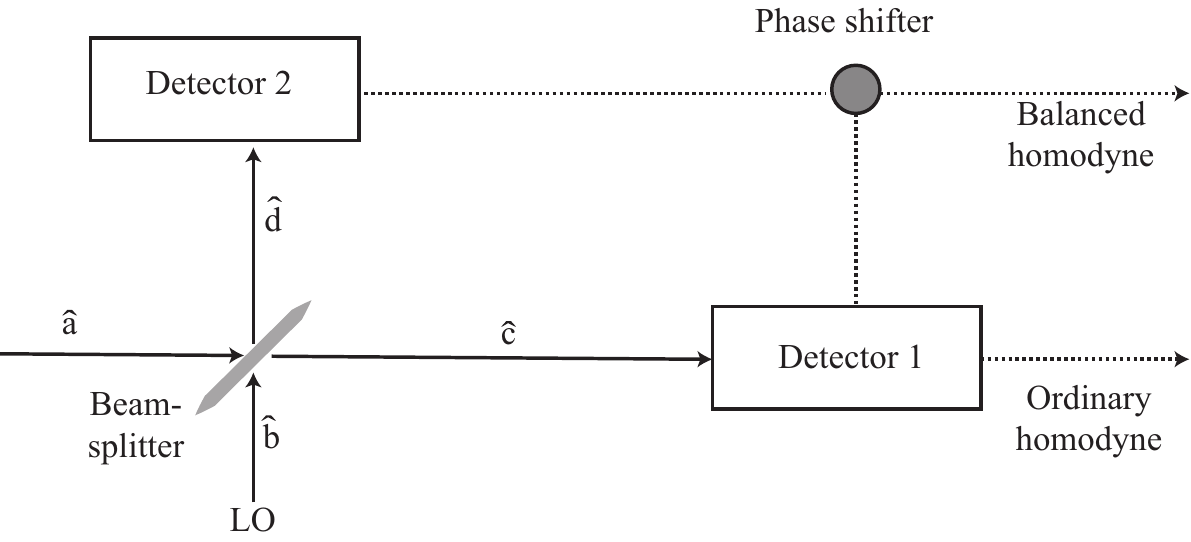}
\caption{\new{Setup for homodyne (and heterodyne) detection.}}
\label{fig:homodyne}
\end{figure}
}

\new{Take the input field (the light leaking from the cavity) $\aop$ and the LO $\hat{b}$. In the beamsplitter, these modes recombine to 
\begin{align}
\hat{c}&=\sqrt{T}\hat{a}+i\sqrt{1-T}\hat{b}\nonumber\\ \hat{d}&=i\sqrt{1-T}\hat{a}+\sqrt{T}\hat{b},
\end{align}
where $T$ is the transmissivity and $1-T$ the reflectivity.
The signals measured in the two detectors are then given by
\begin{align}\label{eq:homsignals}
\hat{c}^\dagger\hat{c}&=T\cop\aop+(1-T)\hat{b}^\dagger\hat{b}+i\sqrt{T(1-T)}(\cop\hat{b}-\hat{b}^\dagger\aop)\nonumber\\
\hat{d}^\dagger\hat{d}&=(1-T)\cop\aop+T\hat{b}^\dagger\hat{b}-i\sqrt{T(1-T)}(\cop\hat{b}-\hat{b}^\dagger\aop).
\end{align}
As one can see, these operators are time-independent if the local oscillator has the same frequency as the input field. Now, use for the LO mode a coherent state $\hat{b}=\beta=\abs{\beta}e^{i\phi_L}$ with large amplitude $\abs{\beta}$ and phase $\phi_L$. With this substitution, the last term in the signal operators \eqref{eq:homsignals} is proportional to
\begin{equation}
\hat{X}_\phi=\frac{1}{2}(\aop e^{-i\phi}+\cop e^{i\phi})
\end{equation}
where $\phi=\phi_L+\pi/2$. $\hat{X}_\phi$ is a quadrature variable, and the most common ones are the conjugate $\hat{X}:=\hat{X}_0$ and $\hat{P}:=\hat{X}_{\pi/2}$.
There are now two possibilities for detection of $\ev{\hat{X}_\phi}$. In the so-called \emph{ordinary homodyne detection} $T\approx1$ and only the photon flux in detector 1 is measured. The input signal term is negligible with respect to the other terms, and the LO-term is constant except from shot noise (and can hence be manually substracted afterwards) so that only the interference term remains. In \emph{balanced homodyne detection}, a beamsplitter with $T=0.5$ is used instead and the signals of both detectors are directly substracted.
Whether ordinary or balanced detection is used has no influence on the quantum trajectory, there only the quadrature variable that is measured is important.
In practical experimental settings, the use of homodyne detection schemes is paramount in a variety of systems. Notable examples are in quantum cryptography \cite{homodynecryptography} and the detection of `spooky action at a distance' \cite{homodynespooky}.
}

\subsection{\new{Heterodyne detection}}
\new{For heterodyne detection, the condition from homodyne detection that the frequency of the LO matches the input field, is relaxed \cite{quantumnoise}. This is equivalent to saying that $\phi$ oscillates at constant frequency. Measurement information on all quadrature variables is encoded in the signal as fourier components \cite{qmac}. Also the corresponding trajectory description of heterodyne detection, at sufficient detuning of the LO, is equivalent to a situation where the light leaving the cavity passes a 50/50 beamsplitter where for half the light a homodyne measurement of $X$ is performed while simultaneously a homodyne measurement of $P$ is performed on the other half of the photocurrent \cite{breuer}.
In an experimental setup, since information on two independent quadratures is retrieved, heterodyne detection is commonly used to obtain phase information. \cite{heterodynephase,heterodyneweitz}. The technique is also used for femtosecond frequency combs \cite{nobel}. 
}

\section{Full equations for \texorpdfstring{$N\Theta$}{NTheta}-Gaussian Trajectories}\label{sec:NTequations}

For the photon-counting unraveling, expectation values evolve as
\begin{align}\label{eq:nthetapcdet}
\partial_t \en &=-\gamma \edndn +2\Im\left[F C_1\right]\\
\partial_t \edndn &=4\Im\left[F C_2\right]-2\Im\left[F C_1\right]\nonumber\\
\partial_t \eph &=\left(\Delta+\frac{U}{2}\right)-U\en-\gamma\edndphsym-\Re\left[F C_3\right]\nonumber\\
\partial_t \edphdph &=-2U\edndphsym-2\Re\left[F C_4\right]+\frac{1}{2}\Im\left[F C_5\right]\nonumber\\
\partial_t \edndphsym &=-U\edndn+2\Im\left[F C_6\right]-\Re\left[F C_7\right]\nonumber\\
\partial_t \unev{1}&=-\gamma \en\unev{1}\nonumber,
\end{align}
where correlators C are given with an expansion in $\en$ in appendix \ref{sec:correlators}. 

Jumps when $\unev{1}=R$ are given by
\begin{align}
\en &\jump \en-1+\frac{\edndn}{\en}\\
\edndn &\jump \edndn\left(1-\frac{\edndn}{\en^2}\right)\nonumber\\
\eph &\jump \eph+\frac{\edndphsym}{\en}\nonumber\\
\edphdph&\jump\edphdph-\frac{\edndphsym^2}{\en^2}+\frac{1}{4\en}\ev{\frac{1}{\nop}}\approx \edphdph+\frac{1}{\en^2}\left(\frac{1}{4}-\edndphsym^2\right)+\frac{\edndn}{4\en^4}\nonumber\\
\edndphsym&\jump\edndphsym\left(1-\frac{\edndn}{\en^2}\right).\nonumber
\end{align}

In the $F=0$ case, evolution \eqref{eq:nthetapcdet} can be solved exactly to
\begin{align}
\en(t)&=\en(0)-\gamma \edndn t\\
\edndn (t)&=\edndn (0)\nonumber\\
\eph (t)&=\eph (0)+\left(\Delta+U(\frac{1}{2}-\en(0))-\gamma \edndphsym(0)\right)t+U\gamma\edndn t^2\nonumber\\
\edphdph(t)&=\edphdph(0)-2U\edndphsym(0)t+U^2\edndn t^2\nonumber\\
\edndphsym(t)&=\edndphsym(0)-U\edndn t\nonumber\\
\unev{1}(t)&=\exp\left(-\gamma\en(0)t+\frac{\gamma^2}{2}\edndn t^2\right).\nonumber
\end{align}
By equating $\unev{1}(t_j)=R$, we find the time to the next jump to be
\begin{equation}
t_j=\gamma^{-1}\frac{\en(0)}{\edndn}\left(1-\sqrt{1+\frac{2\edndn\ln(R)}{\en(0)^2}}\right).
\end{equation}


\vbox{

In the heterodyne unraveling, equations for the evolution are given by

\begin{align}
d\en=&\left[2\Im\left[FC_1\right]-\gamma\en\right]dt\\
&+2\Re\left[(C_2-C_1)\sqrt{\gamma} dZ\right]\nonumber\\
d\edndn &=\left[4\Im\left[FC_2\right]-2\Im\left[FC_1\right]-2\gamma\edndn+\gamma\en\right]dt\nonumber\\
&-2\gamma\abs{C_2-C_1}^2 dt\nonumber\\
&+2\sqrt{\gamma}\Re\left[D_3-2C_2+C_1(1-\edndn)dZ\right]\nonumber\\
d \eph &=\left[\left(\Delta+\frac{U}{2}\right)-U\en-\Re\left[F C_3\right]\right]dt\nonumber\\
&+2\sqrt{\gamma}\Re\left[C_6 dZ\right]\nonumber\\
d\edphdph &=\left[-2U\edndphsym-2\Re\left[F C_4\right]+\frac{1}{2}\Im\left[F C_5\right]+\frac{\gamma}{4}\ev{\nop^{-1}}\right]dt\nonumber\\
&-2\gamma\abs{C_6}^2 dt\nonumber\\
&+2\sqrt{\gamma}\Re\left[(D_1-\edphdph C_1) dZ\right]\nonumber\\
d\edndphsym &=\left[-U\edndn+2\Im\left[F C_6\right]-\Re\left[F C_7\right]-\gamma\edndphsym\right]dt\nonumber\\
&-2\gamma \Re\left[(C_2-C_1)C_6^*\right]dt\nonumber\\
&+2\sqrt{\gamma}\Re\left[\left(-C_6-(\edndphsym+\frac{i}{2})C_1+D_2\right)dZ\right]\nonumber.
\end{align}}

\section{Expansions for correlators of an \texorpdfstring{$N\Theta$}{NTheta}-Gaussian state}\label{sec:correlators}

The expanded correlators are 
\begin{align}\label{eq:correlatoren}
\ev{\cphexp}&=e^{-i\eph-\frac{\edphdph}{2}}\\
C_1:=\ev{\sqrt{\nop}\cphexp}&\approx\sqrt{\en}e^{-i\eph-\frac{\edphdph}{2}}\left(1+\frac{1}{2\en}(\frac{1}{2}-i\edndphsym)-\frac{\edndn}{8\en^2}\right)\nonumber\\
C_2:=\ev{\dn\sqrt{\nop}\cphexp}&\approx\sqrt{\en}e^{-i\eph-\frac{\edphdph}{2}}\left[\frac{1}{2}-i\edndphsym\nonumber\right.\\&\left.+\frac{1}{2\en}\left(\edndn+\frac{1}{4}-\frac{i}{2}\edndphsym-\edndphsym^2\right)\right.\nonumber\\
&\left. +\frac{3}{8\en^2}\left(\frac{-\edndn}{2}+i\edndphsym\edndn\right)
\right]\nonumber\\
C_3:=\ev{\nop^{-\frac{1}{2}}\cphexp}&\approx \frac{e^{-i\eph-\frac{\edphdph}{2}}}{\sqrt{\en}}\left[1+\frac{1}{2\en}\left(\frac{-1}{2}+i\edndphsym\right)+\frac{3\edndn}{8\en^2}\right]\nonumber\\
C_4:=\ev{\nop^{-\frac{1}{2}}\dph\cphexp}&\approx \frac{e^{-i\eph-\frac{\edphdph}{2}}}{\sqrt{\en}}\left[-i\edphdph-\frac{1}{2\en}\left(\edndphsym+\frac{i}{2}\right)\left(1-\edphdph\right)\right.\nonumber\\ &\left.-\frac{3i}{8\en^2}\left(2(\edndphsym+\frac{i}{2})^2+\edndn\edphdph\right)\right]\nonumber\\
C_5:=\ev{\nop^{-\frac{3}{2}}\cphexp}&\approx\frac{e^{-i\eph-\frac{\edphdph}{2}}}{\sqrt{\en}^3}\left[ 1+\frac{1}{2\en}\left(\frac{-3}{2}+3i\edndphsym\right)+\frac{15\edndn}{8\en^2}\right]\nonumber\\
C_6:=\ev{\sqrt{\nop}\dph\cphexp}&\approx \sqrt{\en}e^{-i\eph-\frac{\edphdph}{2}}\left[-i\edphdph+\frac{1}{2\en}\left(\edndphsym+\frac{i}{2}\right)\left(1-\edphdph\right)\right.\nonumber\\ &\left.+\frac{i}{8\en^2}\left(2(\edndphsym+\frac{i}{2})^2+\edndn\edphdph\right)\right]\nonumber\\
C_7:=\ev{\dn\nop^{-\frac{1}{2}}\cphexp}&\approx \sqrt{\en}e^{-i\eph-\frac{\edphdph}{2}}\left[\frac{1}{2\en}\left(1-2i\edndphsym\right)-\frac{\edndn}{2\en^2}\right];\nonumber
\end{align}
\begin{align}
D_1:=\ev{\sqrt{\nop}\dph\dph\cphexp}\approx &\sqrt{\en}e^{-i\eph-\frac{\edphdph}{2}}\left[(1-\edphdph)\edphdph\right.\\&\left.-\frac{i}{2\en}(3\edphdph-\edphdph^2)\left(\edndphsym+\frac{i}{2}\right)\right.\nonumber\\
&\left.-\frac{1}{8\en^2}\left(\edndn\edphdph+2(\edndphsym+\frac{i}{2})^2\right)\right]\nonumber\\
D_2:=\ev{\sqrt{\nop}\dn\dph\cphexp}&\approx\sqrt{\en}e^{-i\eph-\frac{\edphdph}{2}}\left[(\edndphsym+\frac{i}{2})(1-\edphdph)\right.\nonumber\\ &\left.-\frac{i}{2\en}\left(\edndn\edphdph+(\edndphsym+\frac{i}{2})^2(2-\edphdph)\right)\right.\nonumber\\
&\left.-\frac{3}{8\en^2}\edndn(\edndphsym+\frac{i}{2})\right]\nonumber\\
D_3:=\ev{\sqrt{\nop}\dn\dn\cphexp}\approx &\sqrt{\en}e^{-i\eph-\frac{\edphdph}{2}}\left[\edndn-(\edndphsym+\frac{i}{2})^2\right.\nonumber\\
&\left.-\frac{i}{2\en}\left(3\edndn(\edndphsym+\frac{i}{2})-6(\edndphsym+\frac{i}{2})^3\right)-\frac{3\edndn}{8\en^2}\right],\nonumber
\end{align}

where for the expansion up to order $\en^{-1}$ between the brackets all Wick contractions are taken into account (resumming over all orders in $\dph$), and for order $\en^{-2}$ higher order correlators between prefactor and exponential are neglected, similar to \cite{castinexpansions}.



\begin{thebibliography}{-------}
\providecommand{\natexlab}[1]{#1}

\bibitem[Fitzpatrick \em{et~al.}(2017)Fitzpatrick, Sundaresan, Li, Koch, and
  Houck]{photontransition}
Fitzpatrick, M.; Sundaresan, N.M.; Li, A.C.Y.; Koch, J.; Houck, A.A.
\newblock Observation of a Dissipative Phase Transition in a One-Dimensional
  Circuit QED Lattice.
\newblock {\em Phys. Rev. X} {\bf 2017}, {\em 7},~011016.
\newblock
  doi:{\changeurlcolor{black}\href{https://doi.org/10.1103/PhysRevX.7.011016}{\detokenize{10.1103/PhysRevX.7.011016}}}.

\bibitem[Hartmann(2016)]{photonsimulation}
Hartmann, M.J.
\newblock Quantum simulation with interacting photons.
\newblock {\em Journal of Optics} {\bf 2016}, {\em 18},~104005.

\bibitem[Noh and Angelakis(2017)]{lightsimulation}
Noh, C.; Angelakis, D.G.
\newblock Quantum simulations and many-body physics with light.
\newblock {\em Reports on Progress in Physics} {\bf 2017}, {\em 80},~016401.

\bibitem[Amo \em{et~al.}(2009)Amo, Lefr{\`e}re, Pigeon, Adrados, Ciuti,
  Carusotto, Houdr{\'e}, Giacobino, and Bramati]{polaritonsuperfluidity}
Amo, A.; Lefr{\`e}re, J.; Pigeon, S.; Adrados, C.; Ciuti, C.; Carusotto, I.;
  Houdr{\'e}, R.; Giacobino, E.; Bramati, A.
\newblock Superfluidity of polaritons in semiconductor microcavities.
\newblock {\em Nature Physics} {\bf 2009}, {\em 5},~805 EP --.

\bibitem[Lai \em{et~al.}(2007)Lai, Kim, Utsunomiya, Roumpos, Deng, Fraser,
  Byrnes, Recher, Kumada, Fujisawa, and Yamamoto]{zeroandpistates}
Lai, C.W.; Kim, N.Y.; Utsunomiya, S.; Roumpos, G.; Deng, H.; Fraser, M.D.;
  Byrnes, T.; Recher, P.; Kumada, N.; Fujisawa, T.; Yamamoto, Y.
\newblock Coherent zero-state and p-state in an exciton-polariton condensate
  array.
\newblock {\em Nature} {\bf 2007}, {\em 450},~529 EP --.

\bibitem[Jacqmin \em{et~al.}(2014)Jacqmin, Carusotto, Sagnes, Abbarchi,
  Solnyshkov, Malpuech, Galopin, Lema\^{\i}tre, Bloch, and Amo]{honeycomb}
Jacqmin, T.; Carusotto, I.; Sagnes, I.; Abbarchi, M.; Solnyshkov, D.D.;
  Malpuech, G.; Galopin, E.; Lema\^{\i}tre, A.; Bloch, J.; Amo, A.
\newblock Direct Observation of Dirac Cones and a Flatband in a Honeycomb
  Lattice for Polaritons.
\newblock {\em Phys. Rev. Lett.} {\bf 2014}, {\em 112},~116402.
\newblock
  doi:{\changeurlcolor{black}\href{https://doi.org/10.1103/PhysRevLett.112.116402}{\detokenize{10.1103/PhysRevLett.112.116402}}}.

\bibitem[Carmichael(2008)]{carmichaelbook}
Carmichael, H.J.
\newblock {\em Statistical methods in quantum optics 2: Non-classical fields};
  Theoretical and mathematical physics, Springer,  2008.

\bibitem[Breuer and Petruccioni(2002)]{breuer}
Breuer, H.P.; Petruccioni, F.
\newblock {\em The theory of open Quantum Systems}; Oxford university press,
  2002.

\bibitem[Daley(2014)]{daleyreview}
Daley, A.J.
\newblock Quantum trajectories and open many-body quantum systems.
\newblock {\em Advances in Physics} {\bf 2014}, {\em 63},~77--149,
  \href{http://xxx.lanl.gov/abs/https://doi.org/10.1080/00018732.2014.933502}{{\normalfont
  [https://doi.org/10.1080/00018732.2014.933502]}}.
\newblock
  doi:{\changeurlcolor{black}\href{https://doi.org/10.1080/00018732.2014.933502}{\detokenize{10.1080/00018732.2014.933502}}}.

\bibitem[Davies(1969)]{davies1969}
Davies, E.B.
\newblock Quantum stochastic processes.
\newblock {\em Comm. Math. Phys.} {\bf 1969}, {\em 15},~277--304.

\bibitem[Plenio and Knight(1998)]{trajectoryreview}
Plenio, M.B.; Knight, P.L.
\newblock The quantum-jump approach to dissipative dynamics in quantum optics.
\newblock {\em Rev. Mod. Phys.} {\bf 1998}, {\em 70},~101--144.
\newblock
  doi:{\changeurlcolor{black}\href{https://doi.org/10.1103/RevModPhys.70.101}{\detokenize{10.1103/RevModPhys.70.101}}}.

\bibitem[Dalibard \em{et~al.}(1992)Dalibard, Castin, and
  M\o{}lmer]{dalibard1992}
Dalibard, J.; Castin, Y.; M\o{}lmer, K.
\newblock Wave-function approach to dissipative processes in quantum optics.
\newblock {\em Phys. Rev. Lett.} {\bf 1992}, {\em 68},~580--583.
\newblock
  doi:{\changeurlcolor{black}\href{https://doi.org/10.1103/PhysRevLett.68.580}{\detokenize{10.1103/PhysRevLett.68.580}}}.

\bibitem[Dum \em{et~al.}(1992)Dum, Zoller, and Ritsch]{zoller1992}
Dum, R.; Zoller, P.; Ritsch, H.
\newblock Monte Carlo simulation of the atomic master equation for spontaneous
  emission.
\newblock {\em Phys. Rev. A} {\bf 1992}, {\em 45},~4879--4887.
\newblock
  doi:{\changeurlcolor{black}\href{https://doi.org/10.1103/PhysRevA.45.4879}{\detokenize{10.1103/PhysRevA.45.4879}}}.

\bibitem[Carmichael(1993)]{Carmichaelorig}
Carmichael, H.
\newblock {\em An Open Systems Approach to Quantum Optics}; Vol. m18, {\em
  Lecture Notes in Physics}, Springer-Verlag,  1993.

\bibitem[Barchielli and Belavkin(1991)]{barchielli1991}
Barchielli, A.; Belavkin, V.P.
\newblock Measurements continuous in time and a posteriori states in quantum
  mechanics.
\newblock {\em Journal of Physics A: Mathematical and General} {\bf 1991}, {\em
  24},~1495.

\bibitem[Wiseman and Milburn(2009)]{qmac}
Wiseman, H.M.; Milburn, G.J.
\newblock {\em Quantum Measurement and Control}; Cambridge University Press,
  2009.

\bibitem[Prosen(2008)]{prosenexact}
Prosen, T.
\newblock Third quantization: a general method to solve master equations for
  quadratic open Fermi systems.
\newblock {\em New Journal of Physics} {\bf 2008}, {\em 10},~043026.

\bibitem[Sedlmayr \em{et~al.}(2018)Sedlmayr, Fleischhauer, and
  Sirker]{fleishhauer}
Sedlmayr, N.; Fleischhauer, M.; Sirker, J.
\newblock Fate of dynamical phase transitions at finite temperatures and in
  open systems.
\newblock {\em Phys. Rev. B} {\bf 2018}, {\em 97},~045147.
\newblock
  doi:{\changeurlcolor{black}\href{https://doi.org/10.1103/PhysRevB.97.045147}{\detokenize{10.1103/PhysRevB.97.045147}}}.

\bibitem[Kramer(2008)]{tdvprev}
Kramer, P.
\newblock A review of the time-dependent variational principle.
\newblock {\em Journal of Physics: Conference Series} {\bf 2008}, {\em
  99},~012009.

\bibitem[Weimer(2015)]{weimerPRL}
Weimer, H.
\newblock Variational Principle for Steady States of Dissipative Quantum
  Many-Body Systems.
\newblock {\em Phys. Rev. Lett.} {\bf 2015}, {\em 114},~040402.
\newblock
  doi:{\changeurlcolor{black}\href{https://doi.org/10.1103/PhysRevLett.114.040402}{\detokenize{10.1103/PhysRevLett.114.040402}}}.

\bibitem[Overbeck and Weimer(2016)]{varint}
Overbeck, V.R.; Weimer, H.
\newblock Time evolution of open quantum many-body systems.
\newblock {\em Phys. Rev. A} {\bf 2016}, {\em 93},~012106.
\newblock
  doi:{\changeurlcolor{black}\href{https://doi.org/10.1103/PhysRevA.93.012106}{\detokenize{10.1103/PhysRevA.93.012106}}}.

\bibitem[McCutcheon \em{et~al.}(2011)McCutcheon, Dattani, Gauger, Lovett, and
  Nazir]{varDattani}
McCutcheon, D.P.S.; Dattani, N.S.; Gauger, E.M.; Lovett, B.W.; Nazir, A.
\newblock A general approach to quantum dynamics using a variational master
  equation: Application to phonon-damped Rabi rotations in quantum dots.
\newblock {\em Phys. Rev. B} {\bf 2011}, {\em 84},~081305.
\newblock
  doi:{\changeurlcolor{black}\href{https://doi.org/10.1103/PhysRevB.84.081305}{\detokenize{10.1103/PhysRevB.84.081305}}}.

\bibitem[Pollock \em{et~al.}(2013)Pollock, McCutcheon, Lovett, Gauger, and
  Nazir]{multisite}
Pollock, F.A.; McCutcheon, D.P.S.; Lovett, B.W.; Gauger, E.M.; Nazir, A.
\newblock A multi-site variational master equation approach to dissipative
  energy transfer.
\newblock {\em New Journal of Physics} {\bf 2013}, {\em 15},~075018.

\bibitem[Suri \em{et~al.}(2017)Suri, Binder, Muralidharan, and
  Vinjanampathy]{optimisation}
Suri, N.; Binder, F.C.; Muralidharan, B.; Vinjanampathy, S.
\newblock Speeding up Thermalisation via Open Quantum System Variational
  Optimisation,  2017,
  \href{http://xxx.lanl.gov/abs/arXiv:1711.08776}{{\normalfont
  [arXiv:1711.08776]}}.

\bibitem[Diehl \em{et~al.}(2010)Diehl, Tomadin, Micheli, Fazio, and
  Zoller]{gutzdm1}
Diehl, S.; Tomadin, A.; Micheli, A.; Fazio, R.; Zoller, P.
\newblock Dynamical Phase Transitions and Instabilities in Open Atomic
  Many-Body Systems.
\newblock {\em Phys. Rev. Lett.} {\bf 2010}, {\em 105},~015702.
\newblock
  doi:{\changeurlcolor{black}\href{https://doi.org/10.1103/PhysRevLett.105.015702}{\detokenize{10.1103/PhysRevLett.105.015702}}}.

\bibitem[Tomadin \em{et~al.}(2011)Tomadin, Diehl, and Zoller]{gutzdm2}
Tomadin, A.; Diehl, S.; Zoller, P.
\newblock Nonequilibrium phase diagram of a driven and dissipative many-body
  system.
\newblock {\em Phys. Rev. A} {\bf 2011}, {\em 83},~013611.
\newblock
  doi:{\changeurlcolor{black}\href{https://doi.org/10.1103/PhysRevA.83.013611}{\detokenize{10.1103/PhysRevA.83.013611}}}.

\bibitem[Joubert-Doriol and Izmaylov(2015)]{TDVP}
Joubert-Doriol, L.; Izmaylov, A.F.
\newblock Problem-free time-dependent variational principle for open quantum
  systems.
\newblock {\em The Journal of Chemical Physics} {\bf 2015}, {\em 142},~134107,
  \href{http://xxx.lanl.gov/abs/https://doi.org/10.1063/1.4916384}{{\normalfont
  [https://doi.org/10.1063/1.4916384]}}.
\newblock
  doi:{\changeurlcolor{black}\href{https://doi.org/10.1063/1.4916384}{\detokenize{10.1063/1.4916384}}}.

\bibitem[Schr\"oder and Chin(2016)]{matrixprod}
Schr\"oder, F.A.Y.N.; Chin, A.W.
\newblock Simulating open quantum dynamics with time-dependent variational
  matrix product states: Towards microscopic correlation of environment
  dynamics and reduced system evolution.
\newblock {\em Phys. Rev. B} {\bf 2016}, {\em 93},~075105.
\newblock
  doi:{\changeurlcolor{black}\href{https://doi.org/10.1103/PhysRevB.93.075105}{\detokenize{10.1103/PhysRevB.93.075105}}}.

\bibitem[Manzoni \em{et~al.}(2017)Manzoni, Chang, and Douglas]{Manzoni2017}
Manzoni, M.T.; Chang, D.E.; Douglas, J.S.
\newblock Simulating quantum light propagation through atomic ensembles using
  matrix product states.
\newblock {\em Nature Communications} {\bf 2017}, {\em 8},~1743.
\newblock
  doi:{\changeurlcolor{black}\href{https://doi.org/10.1038/s41467-017-01416-4}{\detokenize{10.1038/s41467-017-01416-4}}}.

\bibitem[Mascarenhas \em{et~al.}(2015)Mascarenhas, Flayac, and
  Savona]{savonaMPO}
Mascarenhas, E.; Flayac, H.; Savona, V.
\newblock Matrix-product-operator approach to the nonequilibrium steady state
  of driven-dissipative quantum arrays.
\newblock {\em Phys. Rev. A} {\bf 2015}, {\em 92},~022116.
\newblock
  doi:{\changeurlcolor{black}\href{https://doi.org/10.1103/PhysRevA.92.022116}{\detokenize{10.1103/PhysRevA.92.022116}}}.

\bibitem[Cui \em{et~al.}(2015)Cui, Cirac, and Ba\~nuls]{PhysRevLett.114.220601}
Cui, J.; Cirac, J.I.; Ba\~nuls, M.C.
\newblock Variational Matrix Product Operators for the Steady State of
  Dissipative Quantum Systems.
\newblock {\em Phys. Rev. Lett.} {\bf 2015}, {\em 114},~220601.
\newblock
  doi:{\changeurlcolor{black}\href{https://doi.org/10.1103/PhysRevLett.114.220601}{\detokenize{10.1103/PhysRevLett.114.220601}}}.

\bibitem[Verstraete \em{et~al.}(2004)Verstraete, Garc\'{\i}a-Ripoll, and
  Cirac]{MPSverstraete}
Verstraete, F.; Garc\'{\i}a-Ripoll, J.J.; Cirac, J.I.
\newblock Matrix Product Density Operators: Simulation of Finite-Temperature
  and Dissipative Systems.
\newblock {\em Phys. Rev. Lett.} {\bf 2004}, {\em 93},~207204.
\newblock
  doi:{\changeurlcolor{black}\href{https://doi.org/10.1103/PhysRevLett.93.207204}{\detokenize{10.1103/PhysRevLett.93.207204}}}.

\bibitem[Zwolak and Vidal(2004)]{MPSsuperspace}
Zwolak, M.; Vidal, G.
\newblock Mixed-State Dynamics in One-Dimensional Quantum Lattice Systems: A
  Time-Dependent Superoperator Renormalization Algorithm.
\newblock {\em Phys. Rev. Lett.} {\bf 2004}, {\em 93},~207205.
\newblock
  doi:{\changeurlcolor{black}\href{https://doi.org/10.1103/PhysRevLett.93.207205}{\detokenize{10.1103/PhysRevLett.93.207205}}}.

\bibitem[Casteels and Wouters(2017)]{gutzmcwim1}
Casteels, W.; Wouters, M.
\newblock Optically bistable driven-dissipative Bose-Hubbard dimer: Gutzwiller
  approaches and entanglement.
\newblock {\em Phys. Rev. A} {\bf 2017}, {\em 95},~043833.
\newblock
  doi:{\changeurlcolor{black}\href{https://doi.org/10.1103/PhysRevA.95.043833}{\detokenize{10.1103/PhysRevA.95.043833}}}.

\bibitem[Casteels \em{et~al.}(2017)Casteels, Wilson, and Wouters]{gutzmcWim}
Casteels, W.; Wilson, R.M.; Wouters, M.
\newblock Gutzwiller Monte Carlo approach for a critical dissipative spin
  model,  2017,  \href{http://xxx.lanl.gov/abs/arXiv:1709.00693}{{\normalfont
  [arXiv:1709.00693]}}.

\bibitem[Pichler \em{et~al.}(2013)Pichler, Schachenmayer, Daley, and
  Zoller]{gutzmczoller}
Pichler, H.; Schachenmayer, J.; Daley, A.J.; Zoller, P.
\newblock Heating dynamics of bosonic atoms in a noisy optical lattice.
\newblock {\em Phys. Rev. A} {\bf 2013}, {\em 87},~033606.
\newblock
  doi:{\changeurlcolor{black}\href{https://doi.org/10.1103/PhysRevA.87.033606}{\detokenize{10.1103/PhysRevA.87.033606}}}.

\bibitem[Daley \em{et~al.}(2009)Daley, Taylor, Diehl, Baranov, and
  Zoller]{tdmrg1}
Daley, A.J.; Taylor, J.M.; Diehl, S.; Baranov, M.; Zoller, P.
\newblock Atomic Three-Body Loss as a Dynamical Three-Body Interaction.
\newblock {\em Phys. Rev. Lett.} {\bf 2009}, {\em 102},~040402.
\newblock
  doi:{\changeurlcolor{black}\href{https://doi.org/10.1103/PhysRevLett.102.040402}{\detokenize{10.1103/PhysRevLett.102.040402}}}.

\bibitem[Kantian \em{et~al.}(2009)Kantian, Dalmonte, Diehl, Hofstetter, Zoller,
  and Daley]{tdmrg2}
Kantian, A.; Dalmonte, M.; Diehl, S.; Hofstetter, W.; Zoller, P.; Daley, A.J.
\newblock Atomic Color Superfluid via Three-Body Loss.
\newblock {\em Phys. Rev. Lett.} {\bf 2009}, {\em 103},~240401.
\newblock
  doi:{\changeurlcolor{black}\href{https://doi.org/10.1103/PhysRevLett.103.240401}{\detokenize{10.1103/PhysRevLett.103.240401}}}.

\bibitem[Barmettler and Kollath(2011)]{tdmrg3}
Barmettler, P.; Kollath, C.
\newblock Controllable manipulation and detection of local densities and
  bipartite entanglement in a quantum gas by a dissipative defect.
\newblock {\em Phys. Rev. A} {\bf 2011}, {\em 84},~041606.
\newblock
  doi:{\changeurlcolor{black}\href{https://doi.org/10.1103/PhysRevA.84.041606}{\detokenize{10.1103/PhysRevA.84.041606}}}.

\bibitem[Mazzucchi \em{et~al.}(2016{\natexlab{a}})Mazzucchi, Kozlowski,
  Caballero-Benitez, and Mekhov]{Mekhov1}
Mazzucchi, G.; Kozlowski, W.; Caballero-Benitez, S.F.; Mekhov, I.B.
\newblock Collective dynamics of multimode bosonic systems induced by weak
  quantum measurement.
\newblock {\em New Journal of Physics} {\bf 2016}, {\em 18},~073017.

\bibitem[Mazzucchi \em{et~al.}(2016{\natexlab{b}})Mazzucchi, Caballero-Benitez,
  and Mekhov]{Mekhov2}
Mazzucchi, G.; Caballero-Benitez, S.F.; Mekhov, I.B.
\newblock Quantum measurement-induced antiferromagnetic order and density
  modulations in ultracold Fermi gases in optical lattices.
\newblock {\em Scientific Reports} {\bf 2016}, {\em 6},~31196 EP --.
\newblock Article.

\bibitem[Gardiner and Zoller(2004)]{quantumnoise}
Gardiner, C.; Zoller, P.
\newblock {\em Quantum Noise: A Handbook of Markovian and Non-Markovian Quantum
  Stochastic Methods with Applications to Quantum Optics (Springer Series in
  Synergetics)}; Springer,  2004.

\bibitem[{Lewin, Mathieu} and {Paul, Séverine}(2014)]{hfb}
{Lewin, Mathieu}.; {Paul, Séverine}.
\newblock A numerical perspective on Hartree-Fock-Bogoliubov theory.
\newblock {\em ESAIM: M2AN} {\bf 2014}, {\em 48},~53--86.
\newblock
  doi:{\changeurlcolor{black}\href{https://doi.org/10.1051/m2an/2013094}{\detokenize{10.1051/m2an/2013094}}}.

\bibitem[Bach \em{et~al.}(2016)Bach, Breteaux, Chen, Fröhlich, and
  Sigal]{hfbbosons}
Bach, V.; Breteaux, S.; Chen, T.; Fröhlich, J.; Sigal, I.M.
\newblock The time-dependent Hartree-Fock-Bogoliubov equations for Bosons,
  2016,  \href{http://xxx.lanl.gov/abs/arXiv:1602.05171}{{\normalfont
  [arXiv:1602.05171]}}.

\bibitem[Zhu(2016)]{bdgbook}
Zhu, J.X.
\newblock {\em Bogoliubov-de Gennes Method and Its Applications}; Springer
  International Publishing,  2016.
\newblock
  doi:{\changeurlcolor{black}\href{https://doi.org/10.1007/978-3-319-31314-6}{\detokenize{10.1007/978-3-319-31314-6}}}.

\bibitem[Montina and Castin(2006)]{castinBCSstochastic}
Montina, A.; Castin, Y.
\newblock Exact BCS stochastic schemes for a time-dependent many-body fermionic
  system.
\newblock {\em Phys. Rev. A} {\bf 2006}, {\em 73},~013618.
\newblock
  doi:{\changeurlcolor{black}\href{https://doi.org/10.1103/PhysRevA.73.013618}{\detokenize{10.1103/PhysRevA.73.013618}}}.

\bibitem[Hudson(1974)]{HUDSON1974}
Hudson, R.
\newblock When is the wigner quasi-probability density non-negative?
\newblock {\em Reports on Mathematical Physics} {\bf 1974}, {\em 6},~249 --
  252.
\newblock
  doi:{\changeurlcolor{black}\href{https://doi.org/https://doi.org/10.1016/0034-4877(74)90007-X}{\detokenize{https://doi.org/10.1016/0034-4877(74)90007-X}}}.

\bibitem[Carusotto and Ciuti(2013)]{qfloflight}
Carusotto, I.; Ciuti, C.
\newblock Quantum Fluids of light.
\newblock {\em Reviews of Modern Physics} {\bf 2013}, {\em 85}.

\bibitem[Sinatra \em{et~al.}(2001)Sinatra, Lobo, and Castin]{tworigSAL}
Sinatra, A.; Lobo, C.; Castin, Y.
\newblock Classical-Field Method for Time Dependent Bose-Einstein Condensed
  Gases.
\newblock {\em Phys. Rev. Lett.} {\bf 2001}, {\em 87},~210404.
\newblock
  doi:{\changeurlcolor{black}\href{https://doi.org/10.1103/PhysRevLett.87.210404}{\detokenize{10.1103/PhysRevLett.87.210404}}}.

\bibitem[Sinatra \em{et~al.}(2002)Sinatra, Lobo, and Castin]{tworigSAL2}
Sinatra, A.; Lobo, C.; Castin, Y.
\newblock The truncated Wigner method for Bose-condensed gases: limits of
  validity and applications.
\newblock {\em Journal of Physics B: Atomic, Molecular and Optical Physics}
  {\bf 2002}, {\em 35},~3599.

\bibitem[Polkovnikov(2010)]{POLKOVNIKOV20101790}
Polkovnikov, A.
\newblock Phase space representation of quantum dynamics.
\newblock {\em Annals of Physics} {\bf 2010}, {\em 325},~1790 -- 1852.
\newblock
  doi:{\changeurlcolor{black}\href{https://doi.org/https://doi.org/10.1016/j.aop.2010.02.006}{\detokenize{https://doi.org/10.1016/j.aop.2010.02.006}}}.

\bibitem[Drummond and Opanchuk(2017)]{twcons}
Drummond, P.D.; Opanchuk, B.
\newblock Truncated Wigner dynamics and conservation laws.
\newblock {\em Phys. Rev. A} {\bf 2017}, {\em 96},~043616.
\newblock
  doi:{\changeurlcolor{black}\href{https://doi.org/10.1103/PhysRevA.96.043616}{\detokenize{10.1103/PhysRevA.96.043616}}}.

\bibitem[Drummond and Hardman(1993)]{TWAadd1}
Drummond, P.D.; Hardman, A.D.
\newblock Simulation of Quantum Effects in Raman-Active Waveguides.
\newblock {\em EPL (Europhysics Letters)} {\bf 1993}, {\em 21},~279.

\bibitem[Carter(1995)]{TWAadd2}
Carter, S.J.
\newblock Quantum theory of nonlinear fiber optics: Phase-space
  representations.
\newblock {\em Phys. Rev. A} {\bf 1995}, {\em 51},~3274--3301.
\newblock
  doi:{\changeurlcolor{black}\href{https://doi.org/10.1103/PhysRevA.51.3274}{\detokenize{10.1103/PhysRevA.51.3274}}}.

\bibitem[Steel \em{et~al.}(1998)Steel, Olsen, Plimak, Drummond, Tan, Collett,
  Walls, and Graham]{TWAadd3}
Steel, M.J.; Olsen, M.K.; Plimak, L.I.; Drummond, P.D.; Tan, S.M.; Collett,
  M.J.; Walls, D.F.; Graham, R.
\newblock Dynamical quantum noise in trapped Bose-Einstein condensates.
\newblock {\em Phys. Rev. A} {\bf 1998}, {\em 58},~4824--4835.
\newblock
  doi:{\changeurlcolor{black}\href{https://doi.org/10.1103/PhysRevA.58.4824}{\detokenize{10.1103/PhysRevA.58.4824}}}.

\bibitem[Blakie† \em{et~al.}(2008)Blakie†, Bradley†, Davis, Ballagh, and
  Gardiner]{TWAexperiment1}
Blakie†, P.; Bradley†, A.; Davis, M.; Ballagh, R.; Gardiner, C.
\newblock Dynamics and statistical mechanics of ultra-cold Bose gases using
  c-field techniques.
\newblock {\em Advances in Physics} {\bf 2008}, {\em 57},~363--455,
  \href{http://xxx.lanl.gov/abs/https://doi.org/10.1080/00018730802564254}{{\normalfont
  [https://doi.org/10.1080/00018730802564254]}}.
\newblock
  doi:{\changeurlcolor{black}\href{https://doi.org/10.1080/00018730802564254}{\detokenize{10.1080/00018730802564254}}}.

\bibitem[Hebenstreit(2016)]{hebensteit}
Hebenstreit, F.
\newblock Vortex formation and dynamics in two-dimensional driven-dissipative
  condensates.
\newblock {\em Phys. Rev. A} {\bf 2016}, {\em 94},~063617.
\newblock
  doi:{\changeurlcolor{black}\href{https://doi.org/10.1103/PhysRevA.94.063617}{\detokenize{10.1103/PhysRevA.94.063617}}}.

\bibitem[Van~Regemortel \em{et~al.}(2017)Van~Regemortel, Casteels, Carusotto,
  and Wouters]{belland}
Van~Regemortel, M.; Casteels, W.; Carusotto, I.; Wouters, M.
\newblock Spontaneous Beliaev-Landau scattering out of equilibrium.
\newblock {\em Phys. Rev. A} {\bf 2017}, {\em 96},~053854.
\newblock
  doi:{\changeurlcolor{black}\href{https://doi.org/10.1103/PhysRevA.96.053854}{\detokenize{10.1103/PhysRevA.96.053854}}}.

\bibitem[Drummond and Walls(1979)]{DrummondWalls}
Drummond, P.D.; Walls, D.F.
\newblock Quantum theory of optical bistability. I. Nonlinear polarisability
  model.
\newblock {\em J. Phys. A.: Math. Gen.} {\bf 1979}, {\em 13},~725.

\bibitem[Ferraro \em{et~al.}(2005)Ferraro, Olivares, and Paris]{GaussQInf}
Ferraro, A.; Olivares, S.; Paris, M.
\newblock {\em Gaussian States in Quantum Information}; Napoli Series on
  physics and Astrophysics, Bibliopolis,  2005.
\newblock arXiv:quant-ph/0503237.

\bibitem[Casteels \em{et~al.}(2016)Casteels, Finazzi, Boité, Storme, and
  Ciuti]{wimcorrfties}
Casteels, W.; Finazzi, S.; Boité, A.L.; Storme, F.; Ciuti, C.
\newblock Truncated correlation hierarchy schemes for driven-dissipative
  multimode quantum systems.
\newblock {\em New Journal of Physics} {\bf 2016}, {\em 18},~093007.

\bibitem[Wouters(2012)]{Wouters2012}
Wouters, M.
\newblock Wave-function Monte Carlo method for polariton condensates.
\newblock {\em Phys. Rev. B} {\bf 2012}, {\em 85},~165303.
\newblock
  doi:{\changeurlcolor{black}\href{https://doi.org/10.1103/PhysRevB.85.165303}{\detokenize{10.1103/PhysRevB.85.165303}}}.

\bibitem[Kuznetsov and Kartsev(2017)]{Kuznetsov2017}
Kuznetsov, I.O.; Kartsev, P.F.
\newblock Simulation of equilibrium particle distribution of the Bose gas of
  polaritons using quantum Monte Carlo.
\newblock {\em Journal of Physics: Conference Series} {\bf 2017}, {\em
  941},~012070.

\bibitem[Le~Boit{\'e}(2015)]{aleboite}
Le~Boit{\'e}, A.
\newblock Strongly correlated photons in arrays of nonlinear cavities.
\newblock PhD thesis, Universit{\'e} Paris Diderot-Paris 7,  2015.

\bibitem[Nieto(1993)]{phaseopreview}
Nieto, M.M.
\newblock Quantum phase and quantum phase operators: some physics and some
  history.
\newblock {\em Physica Scripta} {\bf 1993}, {\em 1993},~5.

\bibitem[L.~Pitaevskii(2003)]{BEC}
L.~Pitaevskii, S.S.
\newblock {\em Bose-Einstein condensation}; The International Series of
  Monographs on Physics, Oxford University Press, USA,  2003.

\bibitem[Johnson \em{et~al.}(2017)Johnson, Szigeti, Schemmer, and
  Bouchoule]{dpcorr1dcondensate}
Johnson, A.; Szigeti, S.S.; Schemmer, M.; Bouchoule, I.
\newblock Long-lived nonthermal states realized by atom losses in
  one-dimensional quasicondensates.
\newblock {\em Phys. Rev. A} {\bf 2017}, {\em 96},~013623.
\newblock
  doi:{\changeurlcolor{black}\href{https://doi.org/10.1103/PhysRevA.96.013623}{\detokenize{10.1103/PhysRevA.96.013623}}}.

\bibitem[Hush \em{et~al.}(2012)Hush, Carvalho, and Hope]{NPW1}
Hush, M.R.; Carvalho, A.R.R.; Hope, J.J.
\newblock Number-phase Wigner representation for scalable stochastic
  simulations of controlled quantum systems.
\newblock {\em Phys. Rev. A} {\bf 2012}, {\em 85},~023607.
\newblock
  doi:{\changeurlcolor{black}\href{https://doi.org/10.1103/PhysRevA.85.023607}{\detokenize{10.1103/PhysRevA.85.023607}}}.

\bibitem[Hush \em{et~al.}(2013)Hush, Szigeti, Carvalho, and Hope]{NPW2}
Hush, M.R.; Szigeti, S.S.; Carvalho, A.R.R.; Hope, J.J.
\newblock Controlling spontaneous-emission noise in measurement-based feedback
  cooling of a Bose–Einstein condensate.
\newblock {\em New Journal of Physics} {\bf 2013}, {\em 15},~113060.

\bibitem[Lewis-Swan \em{et~al.}(2016)Lewis-Swan, Olsen, and
  Kheruntsyan]{TWAexperiment2}
Lewis-Swan, R.J.; Olsen, M.K.; Kheruntsyan, K.V.
\newblock Approximate particle number distribution from direct stochastic
  sampling of the Wigner function.
\newblock {\em Phys. Rev. A} {\bf 2016}, {\em 94},~033814.
\newblock
  doi:{\changeurlcolor{black}\href{https://doi.org/10.1103/PhysRevA.94.033814}{\detokenize{10.1103/PhysRevA.94.033814}}}.

\bibitem[M{\o}lmer \em{et~al.}(1993)M{\o}lmer, Castin, and Dalibard]{Molmer:93}
M{\o}lmer, K.; Castin, Y.; Dalibard, J.
\newblock Monte Carlo wave-function method in quantum optics.
\newblock {\em J. Opt. Soc. Am. B} {\bf 1993}, {\em 10},~524--538.
\newblock
  doi:{\changeurlcolor{black}\href{https://doi.org/10.1364/JOSAB.10.000524}{\detokenize{10.1364/JOSAB.10.000524}}}.

\bibitem[Krämer \em{et~al.}(2018)Krämer, Plankensteiner, Ostermann, and
  Ritsch]{qotoolbox}
Krämer, S.; Plankensteiner, D.; Ostermann, L.; Ritsch, H.
\newblock QuantumOptics.jl: A Julia framework for simulating open quantum
  systems.
\newblock {\em Computer Physics Communications} {\bf 2018}, {\em 227},~109 --
  116.
\newblock
  doi:{\changeurlcolor{black}\href{https://doi.org/https://doi.org/10.1016/j.cpc.2018.02.004}{\detokenize{https://doi.org/10.1016/j.cpc.2018.02.004}}}.

\bibitem[Adam \em{et~al.}(2014)Adam, Mechler, Szalay, and
  Koniorczyk]{numberanstates}
Adam, P.; Mechler, M.; Szalay, V.; Koniorczyk, M.
\newblock Intelligent states for a
  number-operator\char21{}annihilation-operator uncertainty relation.
\newblock {\em Phys. Rev. A} {\bf 2014}, {\em 89},~062108.
\newblock
  doi:{\changeurlcolor{black}\href{https://doi.org/10.1103/PhysRevA.89.062108}{\detokenize{10.1103/PhysRevA.89.062108}}}.

\bibitem[Haroche and Raimond(2006)]{haroche2006exploring}
Haroche, S.; Raimond, J.
\newblock {\em Exploring the Quantum: Atoms, Cavities, and Photons}; Oxford
  Graduate Texts, OUP Oxford,  2006.

\bibitem[Brun(2002)]{simplemodel}
Brun, T.A.
\newblock A simple model of quantum trajectories.
\newblock {\em American Journal of Physics} {\bf 2002}, {\em 70},~719–737.
\newblock
  doi:{\changeurlcolor{black}\href{https://doi.org/10.1119/1.1475328}{\detokenize{10.1119/1.1475328}}}.

\bibitem[Walls and Milburn(2008)]{walls_milburn}
Walls, D.F.; Milburn, G.J.
\newblock {\em Quantum Optics}, 2 ed.; Springer.,  2008.

\bibitem[Koczyk \em{et~al.}(1996)Koczyk, Wiewiór, and Radzewicz]{pcexperiment}
Koczyk, P.; Wiewiór, P.; Radzewicz, C.
\newblock Photon counting statistics—Undergraduate experiment.
\newblock {\em American Journal of Physics} {\bf 1996}, {\em 64},~240.

\bibitem[Scully and Zubairy(1997)]{scully_zubairy}
Scully, M.O.; Zubairy, M.S.
\newblock {\em Quantum optics}; Cambridge University Press,  1997.

\bibitem[Xu \em{et~al.}(2009)Xu, Gallion, and Mendieta]{homodynecryptography}
Xu, Q.; Gallion, P.; Mendieta, F.
\newblock Optical homodyne detection and applications in quantum cryptography.
\newblock PhD thesis, T\'el\'ecom ParisTech,  2009.

\bibitem[Fuwa \em{et~al.}(2015)Fuwa, Takeda, Zwierz, Wiseman, and
  Furusawa]{homodynespooky}
Fuwa, M.; Takeda, S.; Zwierz, M.; Wiseman, H.M.; Furusawa, A.
\newblock Experimental proof of nonlocal wavefunction collapse for a single
  particle using homodyne measurements.
\newblock {\em Nature Communications} {\bf 2015}, {\em 6},~6665 EP --.
\newblock Article.

\bibitem[D'Ariano and Paris(1994)]{heterodynephase}
D'Ariano, G.M.; Paris, M.G.A.
\newblock Lower bounds on phase sensitivity in ideal and feasible measurements.
\newblock {\em Phys. Rev. A} {\bf 1994}, {\em 49},~3022--3036.
\newblock
  doi:{\changeurlcolor{black}\href{https://doi.org/10.1103/PhysRevA.49.3022}{\detokenize{10.1103/PhysRevA.49.3022}}}.

\bibitem[Schmitt \em{et~al.}(2016)Schmitt, Damm, Dung, Wahl, Vewinger, Klaers,
  and Weitz]{heterodyneweitz}
Schmitt, J.; Damm, T.; Dung, D.; Wahl, C.; Vewinger, F.; Klaers, J.; Weitz, M.
\newblock Spontaneous Symmetry Breaking and Phase Coherence of a Photon
  Bose-Einstein Condensate Coupled to a Reservoir.
\newblock {\em Phys. Rev. Lett.} {\bf 2016}, {\em 116},~033604.
\newblock
  doi:{\changeurlcolor{black}\href{https://doi.org/10.1103/PhysRevLett.116.033604}{\detokenize{10.1103/PhysRevLett.116.033604}}}.

\bibitem[Hänsch()]{nobel}
Hänsch, T.W.
\newblock Passion for precision.
\newblock
  \url{https://www.nobelprize.org/nobel_prizes/physics/laureates/2005/hansch-lecture.pdf}.

\bibitem[Mora and Castin(2003)]{castinexpansions}
Mora, C.; Castin, Y.
\newblock Extension of Bogoliubov theory to quasicondensates.
\newblock {\em Phys. Rev. A} {\bf 2003}, {\em 67},~053615.
\newblock
  doi:{\changeurlcolor{black}\href{https://doi.org/10.1103/PhysRevA.67.053615}{\detokenize{10.1103/PhysRevA.67.053615}}}.

\end{thebibliography}



\end{document}